\documentclass[aps,prd,nofootinbib,two column]{revtex4-2}
\usepackage{graphicx}
\usepackage{subcaption}
\usepackage{natbib}
\usepackage{mathtools}
\usepackage{slashed}
\usepackage{url}
\usepackage[normalem]{ulem}
\usepackage{enumitem}
\usepackage{mathrsfs}
\usepackage{float}
\usepackage{multirow}
\usepackage{xcolor}
\usepackage{booktabs}
\usepackage[section]{placeins}

\usepackage[colorlinks=true, linkcolor=blue, citecolor=teal, urlcolor=purple]{hyperref}
\newcommand{\orcid}[1]{\href{https://orcid.org/#1}{\includegraphics[width=8pt]
		{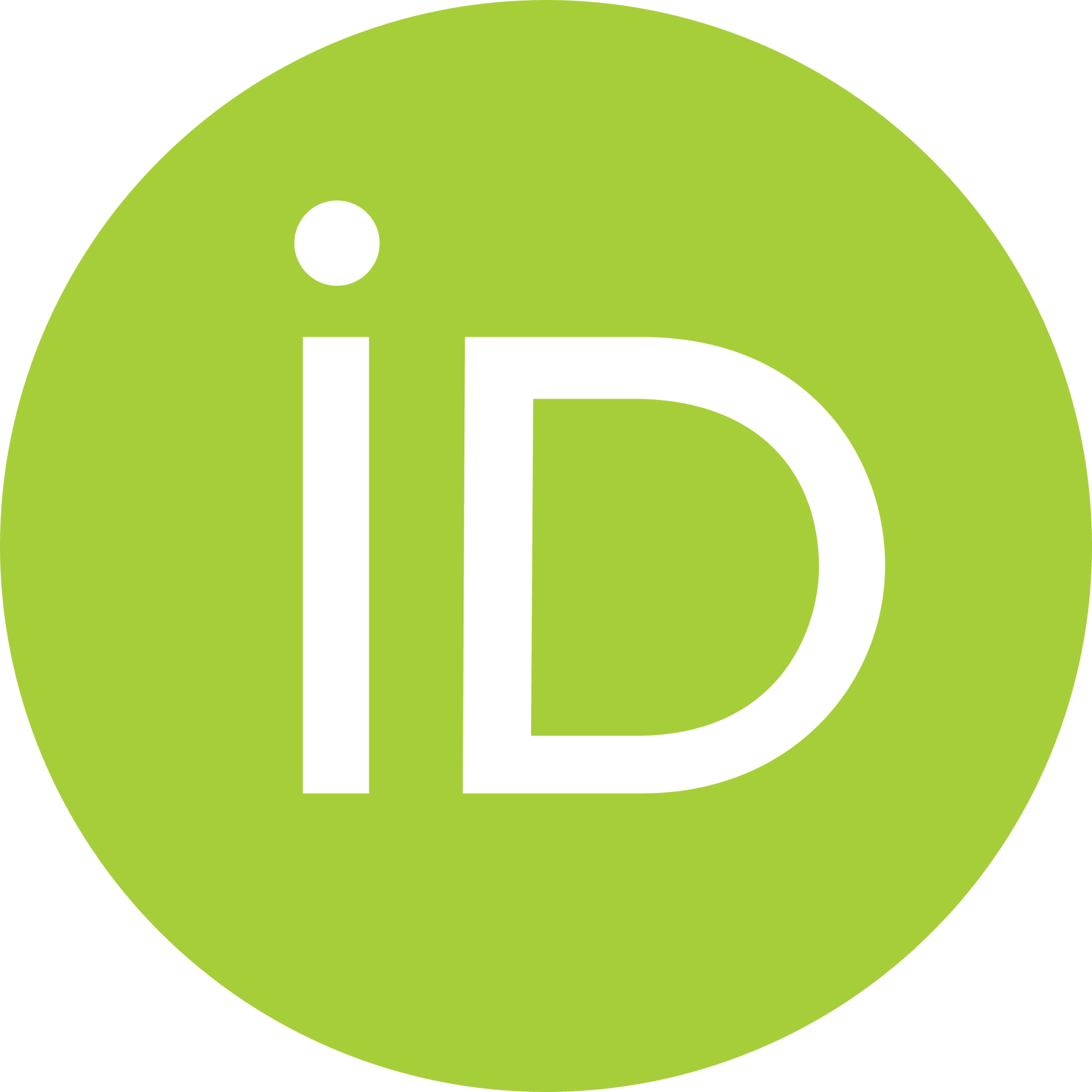}}}

\begin{document}

	\title{Elliptic flow in an expanding and rotating fireball} 

\author{Ashutosh Dwibedi\orcid{https://orcid.org/0009-0004-1568-2806}}
\email{ashutoshdwibedi92@gmail.com}	
\affiliation{Department of Physics, Indian Institute of Technology Bhilai, Kutelabhata, Durg, 491002, Chhattisgarh, India}

\author{Anupam Panja\orcid{https://orcid.org/0009-0005-9692-0868}}
\email{anupampanja55@gmail.com}
\affiliation{Department of Physics, Indian Institute of Technology Bhilai, Kutelabhata, Durg, 491002, Chhattisgarh, India}  
\author{Sabyasachi Ghosh\orcid{https://orcid.org/0000-0003-1212-824X}}
\email{sabya@iitbhilai.ac.in}
\affiliation{Department of Physics, Indian Institute of Technology Bhilai, Kutelabhata, Durg, 491002, Chhattisgarh, India}
	
\begin{abstract}
We investigate the role of initial orbital angular momentum in generating elliptic flow in off-central heavy-ion collisions. The conventional expanding fireball picture is extended for the first time to a simultaneously expanding and rotating fireball characterized by an angular velocity. Spherical and spheroidal emission geometries with two different rotating flow profiles, reminiscent of global and differential rotation, are considered for our investigation. While the transverse-momentum spectra at midrapidity are largely insensitive to the geometry and flow profile for realistic values of angular velocity, the elliptic flow is strongly affected by anisotropic expansion and rotation. A spherical fireball generates elliptic flow solely through the momentum anisotropy induced by rotation, but its magnitude remains below the observed values for realistic angular velocities. In contrast, an anisotropically expanding and rotating spheroidal fireball exhibits a $\sim 10\%$ enhancement of elliptic flow for $\sqrt{s_{\rm NN}}=7.7$ GeV, which reduces to a $2\%$ for $\sqrt{s_{\rm NN}}=39$ GeV relative to its non-rotating counterpart. Within our fireball framework, this indicates that vorticity can contribute at the $2-10\%$ level to the observed elliptic flow.

\end{abstract}

\maketitle
\section{Introduction}\label{Intro}
Nuclei colliding off-center in heavy-ion collisions (HICs) carry substantial orbital angular momentum (OAM) about the collision center, whose magnitude depends on the impact parameter, beam energy, and system size~\cite{Jiang:2016woz,Becattini:2007sr}. For symmetric colliding system with mass number $A$, center-of-mass energy per nucleon $\sqrt{s_{\rm NN}}$, and impact parameter $\tilde{b}$, the initial OAM can be estimated as $L_{0}\sim A\tilde{b}\sqrt{s_{\rm NN}}/2$. For $\sqrt{s_{\rm NN}}=7.7$ GeV, corresponding to the Beam Energy Scan (BES) program at the Relativistic Heavy Ion Collider (RHIC), and $\tilde{b}=7$ fm, this gives $L_{0}\sim2.7\times10^{4}\hbar$. A fraction of the initial OAM is deposited in the reaction zone and transferred to the produced Quark-Gluon Plasma (QGP)~\cite{Harris:2023tti}. The OAM of the produced medium has been proposed to be probed through the polarization of emitted hadrons, arising from spin-orbit (vorticity) coupling~\cite{Liang:2004ph,Gao:2007bc,Betz:2007kg,Becattini:2007sr,Becattini:2007nd,Becattini:2013fla}. Subsequently, the STAR Collaboration measured the global polarization of $\Lambda$ hyperons perpendicular to the reaction plane~\cite{STAR:2017ckg}, followed by many more measurements of local and global polarization of various hyperons and spin alignment of vector mesons~\cite{STAR:2018gyt,STAR:2019erd,STAR:2020xbm,STAR:2021beb,STAR:2023eck,STAR:2023nvo,ALICE:2021pzu,ALICE:2022dyy,STAR:2022fan} supporting the description of vortical QGP. For recent reviews related to the polarization phenomena, see Refs.~\cite{Becattini:2020ngo,Becattini:2021Strongly,Niida:2024ntm,Becattini:2024uha,Dey:2026epy}. In this paper, however, we ask the following questions: whether the initial OAM can create azimuthal anisotropies (specifically the elliptic flow $v_{2}$~\cite{Heinz:2013th,STAR:2013ayu}) in the resulting momentum spectra of the produced hadrons~\cite{STAR:2017sal}? And how does one quantify it in a simple Blast-Wave (BW)~\cite{PhysRevLett.42.880,Bondorf:1978kz,Schnedermann:1993ws,Broniowski:2001we,Broniowski:2001uk,Huovinen:2001cy,Florkowski:2004tn,Retiere:2003kf,Tomasik:2024uuq,Alam:2026ixb} like model? The first question, in principle, has been answered by Becattini \textit{et al.} in Ref.~\cite {Becattini:2007nd} by assuming a globally rotating thermalized system and in Ref.~\cite{Becattini:2007sr} within an intuitive hydrodynamic framework. The physical reason behind the modification in the transverse momentum ($p_{T}=\sqrt{p_{x}^{2}+p_{y}^{2}}$) spectra, as well as elliptic flow, can be attributed to the centrifugal effect, where particles traveling in the direction perpendicular to the OAM, e.g., along the $x$-axis (see Fig.~\eqref{fig:ellipsoid_grad} (a)) receive an extra push, which should reflect in the final hadron spectra and flow of HICs. The momentum spectra~\cite{STAR:2017sal} and elliptic flow~\cite{Heinz:2013th,STAR:2013ayu} of hadrons have previously been studied in frameworks that do not explicitly include the initial OAM. Also, the BW formalism~\cite{PhysRevLett.42.880,Bondorf:1978kz,Schnedermann:1993ws,Broniowski:2001we,Broniowski:2001uk,Huovinen:2001cy,Florkowski:2004tn,Retiere:2003kf,Tomasik:2024uuq,Alam:2026ixb} is designed for the same scenario. Here, we have developed for the first time a BW-like model that incorporates the effect of this initial OAM. We then subsequently evaluated the $p_{T}$ spectra and $v_{2}$ for realistic medium expansion and rotation parameters.  It should be noted that there is a plethora of theoretical studies focusing on the thermodynamics and transport parameters~\cite {Chen:2015hfc,Mameda:2015ria,Jiang:2016wvv,Ebihara:2016fwa,Chernodub_2017,Chernodub:2017ref,Chernodub:2017mvp,Chernodub:2020qah,Wang:2018sur,WeiMingHua:2020eee,Sun:2021hxo,Xu:2022hql,Sun:2023kuu,Fujimoto:2021xix,Mukherjee:2023ijv,Mukherjee:2023qvq,Pradhan:2023rvf,Sahoo:2023xnu,Aung:2023pjf,Dwibedi:2023akm,Dwibedi:2024amt,Padhan:2024edf,Pradhan:2025pol,Sahoo:2025fif,Padhan:2025qhz,Dwibedi:2025boz,Sahoo:2026lrw,Padhan:2026mwg,Kumar:2026dtz,Padhan:2026oyt} of globally rotating QGP motivated by this initial OAM. 

Relativistic hydrodynamics~\cite{Kolb:2003dz,Heinz:2013th,Gale:2013da,Romatschke:2017ejr,De:2022yxq,Ali:2024zvp,10.1093/ptep/pts014}, as well as transport models~\cite{Bass:1998ca,Lin:2004en,SMASH:2016zqf,Cassing:2009vt}, have been quite successful in describing the space-time evolution of matter produced in HICs. More specifically, the formation of vorticity\footnote{It should be noted that while a rotating fluid necessarily carries OAM and vorticity, $\vec{\omega}\equiv\frac{1}{2}(\vec{\nabla}\times\vec{v})$, the reverse is not necessarily true.} in HICs has been investigated extensively within hydrodynamic~\cite{Csernai:2013bqa,Becattini:2013vja,Karpenko:2016jyx,Becattini:2015ska,Karpenko:2021wdm} and transport approaches~\cite{Deng:2016gyh,Jiang:2016woz,Huang:2020dtn}. While these numerical simulations provide a detailed characterization of the system and reproduce experimental observations with good accuracy, the success of BW models in describing momentum spectra~\cite{Schnedermann:1993ws,Broniowski:2001we,Broniowski:2001uk,Huovinen:2001cy,Florkowski:2004tn,Retiere:2003kf,Tomasik:2024uuq,Alam:2026ixb} and, more recently, spin polarization~\cite{Florkowski:2019voj,Florkowski:2021xvy,Banerjee:2024xnd} should not be overlooked. BW models provide a transparent and time-efficient description of the system at kinetic freeze-out. They are therefore widely used in the literature.

In this work, we generalize the traditional BW framework~\cite{Schnedermann:1993ws} by incorporating a rotating component of the fluid velocity into the standard expanding flow. The rotational component is introduced in accordance with the OAM carried by the matter in off-central HICs. We consider two fireball geometries, namely, spherical and spheroidal, to disentangle the effects of geometry, expansion, and rotation on the $p_T$ spectra and elliptic flow $v_2$. Two different flow profiles are considered in each of these geometries: spheroid I and II, and sphere I and II. The second profiles are characterized by uniform rotation along with expansion, whereas in the first profiles, differential rotations are assumed where the angular velocities are position dependent. To parameterize the time dependence of the fireball dimensions from which the expansion velocities are derived, we employ the formalism used in Ref.~\cite{Gossiaux:2011ea} and in our previous works~\cite{Rai:2026dda,Dwibedi:2026yhl}. Finally, we use realistic values of the angular velocity at kinetic freeze-out obtained from transport simulations~\cite{Jiang:2016woz}, where the angular velocity is estimated from the volume-averaged local vorticity. These values are then used to calculate the $p_T$-spectra and $v_2$ of the pion. The observed $p_T$-spectra show little sensitivity if the freeze-out values of the angular velocity are used. However, one can see a clear enhancement of the $p_{T}$ spectra at high $p_{T}$ by using a higher value of angular velocity. On the other hand, the elliptic flows are significantly enhanced by the magnitude of the angular velocity at kinetic freeze-out. Since the angular velocity at low $\sqrt{s_{\rm NN}}$ has been predicted to be higher~\cite{Jiang:2016woz}, we see a higher increase in the elliptic flow at lower $\sqrt{s_{\rm NN}}$ as a function of $p_{T}$.
We also provide the percentage enhancement in the $p_{T}$ averaged elliptic flow due to the effect of rotation for identified hadrons at RHIC BES energies $\sqrt{s_{\rm NN}}=7.7$--$39$ GeV.
 
The remainder of this paper is organized as follows. Sec.~\eqref{offcentral-OAM} provides a brief introduction to off-central HICs and the OAM deposited in the reaction zone, and motivates the generalization of traditional expanding fireball models to simultaneously expanding and rotating fireballs. Sec.~\eqref{fireball_rot} introduces the fireball geometries and flow profiles. In Sec.~\eqref{results}, we present the $p_T$ spectra and $v_2$ obtained using both spherical and spheroidal geometries with two distinct flow profiles. Sec.~\eqref{summary} summarizes our main findings. Throughout this paper, we use natural units $(c=\hbar=k_B=1)$ and adopt the mostly-minus metric convention, $g_{\mu\nu}=\mathrm{diag}(1,-1,-1,-1)$.

\section{Model Description}\label{modeldesc}
    \subsection{Off-central collisions and orbital angular momentum}\label{offcentral-OAM}
    \begin{figure*}[t]
		\centering
		\begin{subfigure}{0.40\textwidth}
			\centering
			\includegraphics[width=\linewidth]{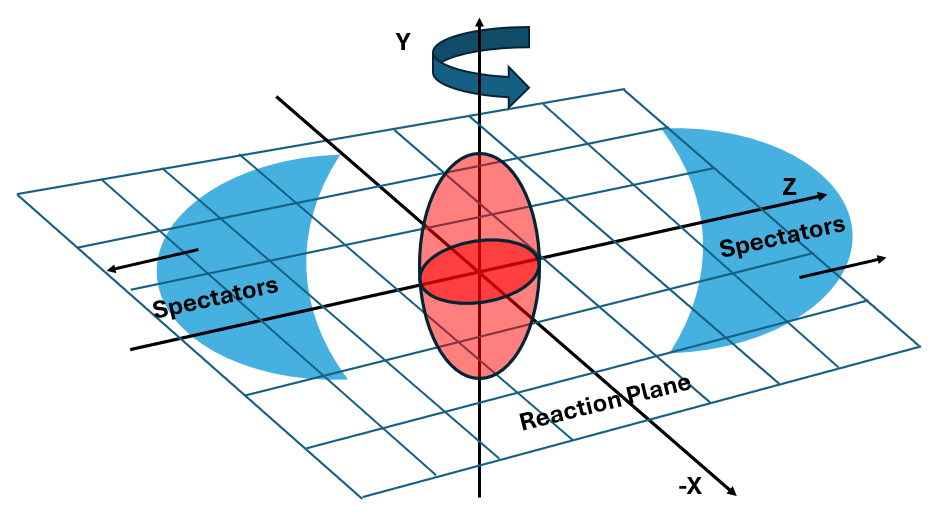}
		\end{subfigure}
        \hfill
        \begin{subfigure}{0.27 \textwidth}
			\centering
			\includegraphics[width=\linewidth]{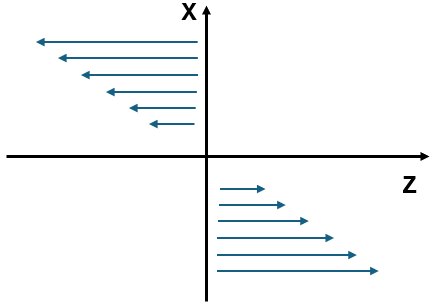}
		\end{subfigure}
        \hfill
        \begin{subfigure}{0.24\textwidth}
			\centering
			\includegraphics[width=\linewidth]{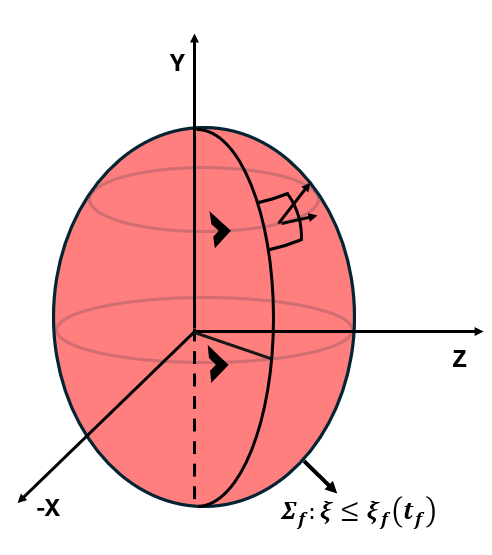}
		\end{subfigure}
		\caption{(a) Geometry of off-central nuclear collisions showing the transferred angular momentum to the participants (left), (b) the preferred initial velocity profile along $z$-axis (middle), and (c) spheroidal expanding and rotating system produced in HICs.}
		\label{fig:ellipsoid_grad}
	\end{figure*}
We begin by briefly recapitulating some relevant aspects of HICs. Shortly after the collision, once the produced matter has thermalized, it is commonly assumed that the matter initially has no velocity in the directions transverse to the beam axis, i.e., $v^x(t=0)=v^y(t=0)=0$. Along the beam direction, however, the matter undergoes longitudinal expansion, which is conventionally described by the Bjorken profile, $v^z=\frac{z}{t}$. While this velocity profile provides a reasonable description of the longitudinal expansion in central collisions, it does not naturally account for angular-momentum conservation in off-central HICs, as emphasized in Ref.~\cite{Becattini:2007sr}.In a non-central collision, as discussed in Sec.~\eqref{Intro}, a fraction of the OAM of the colliding nuclei is deposited in the reaction zone owing to the inhomogeneity of the participant density in the transverse plane~\cite{Jiang:2016woz,Becattini:2007sr}. The remaining OAM is carried away by the spectator nuclei, as depicted in Fig.~\eqref{fig:ellipsoid_grad}(a). The fraction of initial OAM deposited in the reaction zone therefore induces an asymmetric longitudinal velocity profile, $v_0^{z}(x)$, as illustrated in Fig.~\eqref{fig:ellipsoid_grad}(b). To make this connection explicit, let us write the total OAM carried by an ideal fluid as~\cite{Becattini:2007sr},
\begin{eqnarray}
\vec{L}(t)&=&\int \vec{r}\times \vec{\mathcal{P}}~ dV, \label{AD1}  
\end{eqnarray}
where $\vec{\mathcal{P}}$ denotes the momentum density of the fluid. The above integral is evaluated over the system volume $V$ at a given time $t$. For an ideal fluid, the momentum density is given by $\mathcal{P}^{i}=T^{0i}$, where the energy-momentum tensor takes the form $T^{\mu\nu}=(\mathcal{E}+P)u^{\mu}u^{\nu}-P\eta^{\mu\nu}$, with $\mathcal{E}$, $P$, and $u^{\mu}=\gamma(1,\vec{v})$ denoting the proper energy density, pressure, and four-velocity of the fluid, respectively. In off-central HICs, the initial OAM is predominantly perpendicular to the reaction plane, i.e., along the $y$-axis. Substituting the ideal-fluid energy-momentum tensor into the expression for the OAM, its $y$-component can be written as
\begin{eqnarray}
 L^{y}(t)&=&\int (zv^{x}-xv^{z})~\gamma^{2}(\mathcal{E}+P)~dV~,\label{AD2}  
\end{eqnarray}
where $\gamma=1/\sqrt{1-\vec{v}^{2}}$. Eq.~\eqref{AD2} has an intuitive interpretation owing to its similarity to the familiar expression for angular momentum in non-relativistic mechanics. In a relativistic fluid, the mass density is effectively replaced by the proper enthalpy density, $h\equiv \mathcal{E}+P$,
which is the enthalpy density measured in the local rest frame of the fluid element. At the initial time $(t=0)$, Eq.~\eqref{AD2} simplifies to
\begin{eqnarray}
L^{y}(t=0)&=&-\int xv_{0}^{z}~\gamma_{0}^{2}(\mathcal{E}_{0}+P_{0})~dV~,\label{AD3}  
\end{eqnarray}
where $\gamma_{0}=[1-(v^{z}_{0})^{2}]^{-1/2}$, and, as usual, we assume the transverse velocities to vanish. It follows from Eq.~\eqref{AD3} that a non-vanishing angular momentum along the $y$-axis requires either the enthalpy density or the velocity profile to be asymmetric in $x$; otherwise, the integrand is odd in $x$ and the integral vanishes. To motivate the initial velocity profile shown in Fig.~\eqref{fig:ellipsoid_grad}(b), Ref.~\cite{Becattini:2007sr} argued that it is more natural to consider a symmetric enthalpy density profile together with an asymmetric longitudinal velocity profile $v^{z}_{0}(x)$. Indeed, if the initial enthalpy density is predominantly determined by the spatial distribution of the participant nucleons, it is reasonable to assume, to a good approximation, that $h_{0}(-x)=h_{0}(x)$. Consequently, a non-vanishing initial OAM requires an asymmetric longitudinal velocity profile, $v^{z}_{0}(-x)\neq v^{z}_{0}(x)$, thereby providing a natural connection between the initial velocity profile and OAM conservation. This asymmetric initial velocity profile can, through the subsequent fluid evolution, generate an additional momentum anisotropy in the transverse $XY$-plane.\footnote{This anisotropy is in addition to that generated by the pressure gradients in off-central HICs.} In particular, fluid elements moving along the $x$-direction can acquire a larger transverse velocity than those moving along the $y$-direction, thereby providing an additional, vorticity-induced mechanism for the generation of elliptic flow, $v_{2}$~\cite{Becattini:2007sr}.

The initially deposited OAM must be preserved throughout the subsequent fluid evolution. Therefore, for off-central HICs, it is essential to construct a flow profile that retains the initial OAM while simultaneously providing a successful description of the spectra and flow observables of the produced particles. The traditional BW models~\cite{Huovinen:2001cy}, however, do not carry a net OAM, as can be readily verified by substituting their flow profile,
\begin{eqnarray}
u^{\mu}=\gamma_{T}\cosh\eta_{L}\left(1,\frac{\vec{\tilde{v}}_{T}}{\cosh\eta_{L}}, \tanh\eta_{L}\right)~,\label{AD4}
\end{eqnarray}
into Eq.~\eqref{AD2}. We emphasize that Eq.~\eqref{AD4} contains an inherent cylindrical symmetry. The longitudinal fluid rapidity is commonly taken to follow the Bjorken form, $\tanh\eta_{L}=z/t$, while the transverse velocities depend only on the transverse coordinates, $\vec{\tilde{v}}_{T}=\tilde{v}_{\rho}(\rho,\phi)\hat{\rho}+\tilde{v}_{\phi}(\rho,\phi)\hat{\phi}$, where $\rho$ and $\phi$, respectively, defines the radius and azimuth on the $XY$-plane\footnote{We generically use a tilde to denote a velocity scaled by a rapidity. Likewise, $\theta$ and $\phi$ are used generically for angular variables. Their precise definitions and conventions are stated in the text for each geometry.}. Similarly, it is easy to see that in other geometric descriptions, such as spherical~\cite{Harabasz:2020sei,Dwibedi:2026yhl} and spheroidal~\cite{Harabasz:2022rdt} fireball models, the net OAM vanishes. These profiles can be physically understood as descriptions of purely expanding flow, without incorporating the rotational component associated with the initial OAM. Although such expanding profiles can successfully describe the radial and elliptic flow of the produced particles, they do not physically capture the scenario in off-central HICs as far as OAM is concerned. In Sec.~\eqref{fireball_rot}, we modify these flow profiles to incorporate the initial OAM by explicitly adding a rotating component to the expanding flow. In Sec.~\eqref{results}, we show that these profiles remain consistent with the experimentally measured $p_T$ spectra and, in addition, generate a vorticity (rotation)-induced contribution to elliptic flow, as first pointed out in Ref.~\cite{Becattini:2007sr}.

\subsection{Rotating fireball models}\label{fireball_rot}
We now turn to providing a more realistic description of the reaction zone via modeling it as a simultaneously expanding and rotating spheroid as shown in Fig.~\eqref{fig:ellipsoid_grad} (c). The spheroidal coordinate system is obtained by revolving the elliptic coordinate system around the $y$ axis.
We use the following parameterization: $x = \lambda \sinh\xi \cos\theta \sin\phi, $ $ y = \lambda \cosh\xi \sin\theta,$ and  $z = \lambda \sinh\xi \cos\theta \cos\phi$, where $
	-\frac{\pi}{2}\leq\theta<\frac{\pi}{2},~ 0\leq\phi<2\pi, \text{ and }  \xi\geq 0$. $\lambda$ is the focal length of the ellipse and is fixed for a given coordinate system. We measure the polar angle $\theta$ from the $x$-axis, whereas the azimuthal angle $\phi$ is measured from the $z$-axis. For a constant $\theta$, one obtains a hyperboloid of revolution by rotating the hyperbola (say, lying on the $XY$ plane) around the $y$-axis. Whereas, for constant $\xi$ and $\phi$, one obtains spheroids and planes with their normal lying in the $ZX$-plane. The major and minor axes of the ellipse is given by, $a\equiv\lambda\cosh\xi$ and $b\equiv\lambda\sinh\xi$, respectively. Similar to the spherically expanding fireball we write two distinct fluid profiles by exploiting the symmetry of the system.
    \begin{equation}
     \text{Spheroid-I}:u^{\mu}=\gamma_{\Omega}\cosh\eta_y\left(1,\frac{\tilde{v}^x}{\cosh\eta_y},v^{y},\frac{\tilde{v}^z}{\cosh\eta_y}\right)\label{AD15}~.
 \end{equation}
We have $\tilde{v}^x=\tilde{v}^x_{0}+\tilde{v}^{x}_{\Omega}, v^{y}=\tanh\eta_{y}, \text{and }  \tilde{v}^z=\tilde{v}^z_{0}+\tilde{v}^{z}_{\Omega}$
 where we split the $x$ and $z$ components of velocity into the expanding and rotating parts, respectively. Note that we should assume the same expansion velocity along $x$ and $z$ direction for the shape of the spheroid to be preserved. The exact form of the components is expressed as,
  \begin{eqnarray}   
   && \tilde{v}^x_{0}=\frac{r}{r_B}v_{b}\cos\theta\sin\phi,~\tilde{v}^{x}_{\Omega}=\Omega  \lambda\sinh\xi\cos\theta\cos\phi,\nonumber\\
   && \tilde{v}^z_{0}=\frac{r}{r_B}v_{b}\cos\theta\cos\phi,~\tilde{v}^{z}_{\Omega}=-\Omega \lambda\sinh\xi\cos\theta\sin\phi,  \nonumber\\
 &&\text{and }v^{y}=\frac{r}{r_{B}}v_{a}\sin\theta~,\label{AD16}
  \end{eqnarray}
  where $v_{a}$ and $v_{b}$ are the two velocity parameters determining the expansion of the spheroid along the major and minor axes, respectively.
  The boundary of the spheroidal fireball is defined by, $0<\xi\leq \xi_{\rm max}=\xi_{\rm B}$. We observe that the angular velocity in this profile is dependent on the point $(\xi,\theta)$ on the $XY$-plane and given by $\Omega_{I}=\Omega/(\gamma_{y}(\xi,\theta)$), where $\gamma_{y}= \cosh\eta_{y}=1/\sqrt{1-(v^{y})^{2}}$. The second profile we assume is similar to the first profile, but now with uniform angular velocity $\Omega_{II}=\Omega$ and no explicit boost scaling for velocities in the $XZ$-plane,
\begin{eqnarray}
     \text{Spheroid-II}: u^{\mu}&=&\gamma\left(1,v_{0}^{x}+v_{\Omega}^{x},v^{y},v_{0}^{z}+v_{\Omega}^{z}\right)\label{AD20}.
\end{eqnarray}
Ignoring the tildes on the left-hand sides of Eq.~\eqref{AD16}, the form of the expanding and rotating part of the velocities for profile-II can be chosen exactly by the right-hand sides of Eq.~\eqref{AD16}.
The expansion of the boundary of the spheroid is due to the increase of its major and minor axes with time in the following manner,
\begin{align}
	a(t) &= a_0
	+ v_\infty\left[t - \frac{1 - e^{-A t}}{A}\right]
	- \Delta v\left[t - \frac{1 - e^{-B t}}{B}\right],
	\label{AD22} \\
	b(t) &= b_0
	+ v_\infty\left[t - \frac{1 - e^{-A t}}{A}\right]
	+ \Delta v\left[t - \frac{1 - e^{-B t}}{B}\right]~.
	\label{AD23}
\end{align}
This hydrodynamics inspired expansion rate of the ellipsoid axes are taken from Ref.~\cite{Gossiaux:2011ea}. The $A$, and  $B$ defines rate at which the average radial flow (associated with asymptotic expansion velocity $v_{\infty}$) and the asymmetric flow (associated with $\Delta v$) is generated. It is important to note that the asymmetric flow parametrized by the quantities $B$ and $\Delta v$ are generated here due to the difference in the pressure gradient in the $XY$ plane and not due to the rotation as discussed in Sec.~\eqref{offcentral-OAM}. 
The net  OAM stored in these profiles can be readily obtained by substituting Eqs.~\eqref{AD16} and \eqref{AD20} in Eq.~\eqref{AD1}. We get a zero OAM in $x$ and $z$ directions and a finite OAM along $y$. For spheroid-I,
\begin{equation}
L^{y}(t)= \lambda^{2}\int \sinh^{2}\xi \cos^{2}\theta  \frac{\Omega}{\cosh\eta_{y}}~h\gamma_{\Omega}^{2} \cosh^{2}\eta_{y} ~dV~.
\label{AD28}
\end{equation}
 Similarly, for the profile Spheroid-II, the $y$-component of the OAM is given by,
\begin{equation}
  L^{y}(t)= \lambda^{2}\int \sinh^{2}\xi \cos^{2}\theta  ~\Omega~h\gamma^{2} dV~. \label{AD30}
\end{equation}
The Eqs.~\eqref{AD28} and \eqref{AD30} have a simple interpretation: it corresponds to the net OAM of a relativistic quasi-rigid (rigid) spheroidal rotor with differential (constant) angular velocity $\Omega/\cosh\eta_{y}$ ($\Omega$). The expansion profile does not yield any OAM, as expected. Apart from the angular velocity, the other terms inside the integrals presented in Eqs.~\eqref{AD28} and \eqref{AD30} may be identified with the moment of inertia of the rotating spheroid.

We now introduce an expanding and rotating spherical fireball. It helps us to disentangle the effect of anisotropic expansion and rotation in generating $v_{2}$.  We denote the models as Sphere-I and Sphere-II, where in the first profile, the angular velocity is dependent on the radius at which the spherical shell is located, and in the other profile, it is uniform.
\begin{eqnarray}
&\text{Sphere-I: } u^{\mu}
 =\gamma_{\Omega}\cosh\eta_{r} \left(1,\tanh\eta_{r}\hat{r},\frac{\Omega r \sin\theta}{\cosh\eta_{r}}\hat{\phi}\right),\label{AD9}\\
& \text{Sphere-II: } u^{\mu}
 =\gamma \left(1,\tanh\eta_{r}\hat{r},\Omega r \sin\theta\hat{\phi}\right)~,\label{AD10}
\end{eqnarray}
where the polar angle $\theta$ and azimuthal angle $\phi$ in the position space are measured from the positive $y$ and $z$ axes, in clockwise and counterclockwise directions, respectively. For Sphere-II, $\Omega_{II}=\Omega$ whereas for, Sphere-I, the angular velocity is radius-dependent $\Omega_{I}(r)=\Omega/\cosh\eta_{r}$. The expansion velocity can be expressed as
\begin{eqnarray}
    v_{r}=\frac{r}{r_{B}} v_{B}~,\label{AD11}
\end{eqnarray}
where $v_{B}$ and $r_{B}$ are, respectively, the velocity and radius corresponding to the boundary of the spherical fireball. The boundary of the sphere is assumed to increase in accordance with Eq.~\eqref{AD22} by keeping $\Delta v=0$, now written for the sphere radius.

The OAM stored in the two profiles can be obtained by substituting the expressions~\eqref{AD9} and \eqref{AD10} in Eq.~\eqref{AD1},
\begin{eqnarray}
\text{Sphere-I: } \vec{L}(t)&=&\hat{j}\int h\gamma^2 \frac{\Omega r^{2}\sin^{2}\theta}{\cosh\eta_{r}}~dV~,\label{AD13}\\
 \text{Sphere-II: }  \vec{L}(t)&=&\hat{j}\int h\gamma^2 ~\Omega r^{2}\sin^{2}\theta~dV\label{AD14}~,
\end{eqnarray}
where we observe that, due to symmetry, the $\phi$ integral over the full period vanishes, and one has only the OAM along the $y$-axis. Like the results of the spheroidal system, the results in Eq.~\eqref{AD13} and Eq.~\eqref{AD14} carry a simple meaning where OAM is given by the moment of inertia times the angular velocity.

We end this section by emphasizing that profile-II of the spherical model is, in fact, a special case of profile-II of the spheroidal model. This equivalence is obtained by adopting the same convention for the polar angle in the two coordinate systems and taking the zero-focal-length limit. Profile-I, however, is intrinsically different in the two geometries. In the spherical case, the azimuthal components of the velocity are scaled by a radial boost, whereas in the spheroidal case, the velocities in the $XZ$-plane are scaled by a boost along the $y$-direction. Consequently, profile-I of the spherical model cannot be obtained as a limiting case of the corresponding spheroidal profile-I.

\section{RESULTS AND DISCUSSION}\label{results}
	\begin{figure*}[t!]
		\centering
			\includegraphics[width=\linewidth]{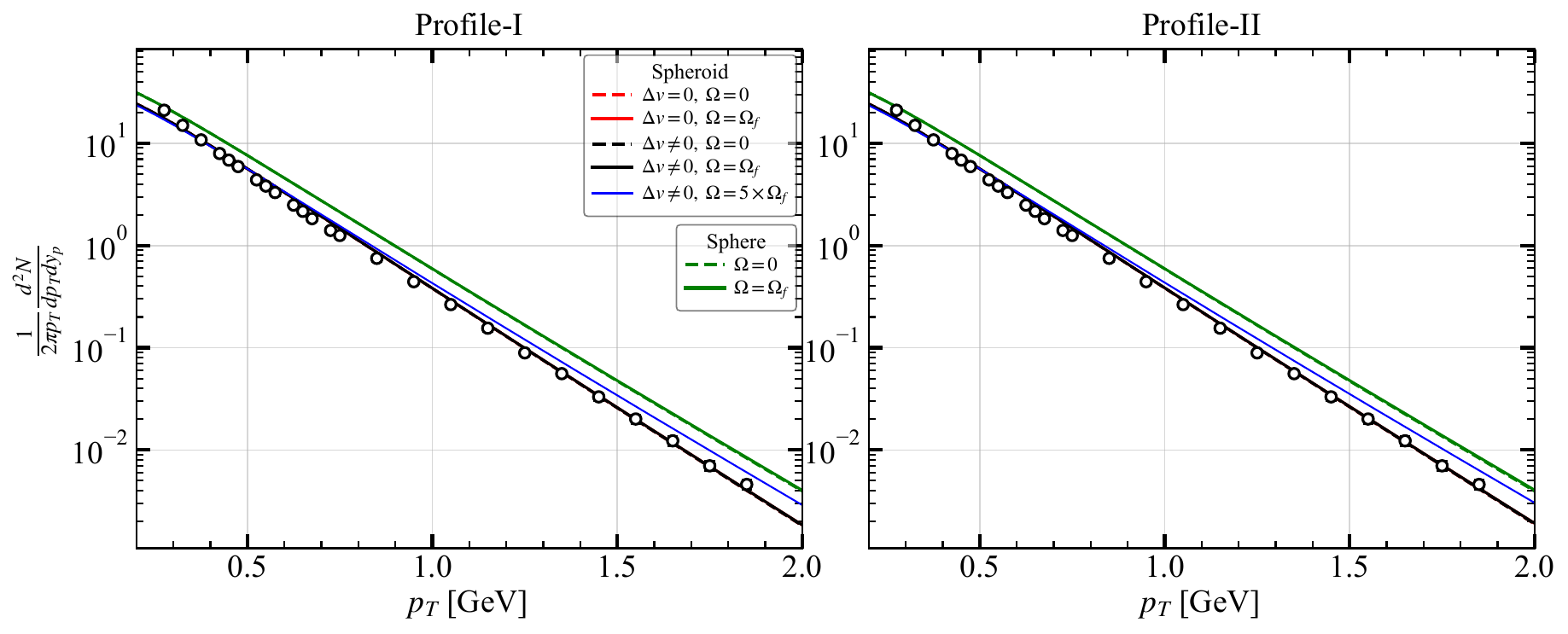}
		\caption{(Color online) Calculated $p_T$ spectra of $\pi^+$ at mid-rapidity ($y_p=0$) for four different scenarios, using profile-I (left) and profile-II (right), compared with the STAR data~\cite{STAR:2017sal} (scattered dots) for $\sqrt{s_{\rm NN}}=7.7$ GeV and centrality class $40-50\%$. Both plots share the same legend.}
		\label{fig:pion_pT}
	\end{figure*}
Having specified the details of both the spherical and spheroidal fireball models, we move to describe the soft observables (low-$p_{T}$ regime of the identified hadrons) of HICs. It is noteworthy that, in HICs, experimentalists finally measure the momentum spectra obtained in the collision and try to characterize the history of the collisions and the produced matter by analyzing the spectra and their anisotropy in different momentum bins. Theoretically, this spectrum is customarily obtained by the prescription of Cooper--Frye~\cite{Cooper:1974mv}, which gives the invariant momentum distribution of the identified hadrons as,
\begin{equation}
	E\frac{d^3N}{d^3\vec{p}}=\frac{g}{(2\pi)^3} \int_{\Sigma} f(x,p)~ p^\mu  d\Sigma_\mu,
	\label{eq:cooper_frye1}
\end{equation}
where $f$ is the single particle distribution function, $g$ is the associated spin degeneracy factor, and $\Sigma$ denotes the kinetic freeze-out hypersurface. The Eq.~\eqref{eq:cooper_frye1} is interpreted as the number of particle worldliness crossing the hypersurface $\Sigma_{\mu}$ per unit momentum space volume $\frac{d^{3}\vec{p}}{E}$. We choose a instantaneous lab time freeze-out at time $t=t_{f}$ which corresponds to $d\Sigma_\mu = (d^3x, \mathbf{0})$ and use the local thermal the Bose--Einstein or Fermi--Dirac distribution,  $f(x,p) =\left[\exp\!\left(\dfrac{u_{\mu}p^{\mu} - \mu}{T}\right) \mp 1\right]^{-1}$. The elliptic flow is obtained as the second Fourier coefficient of the expansion of the spectra in the azimuthal angle $\phi_{p}$ measured in $XY$-plane~\cite{Kolb:2003dz}.
The expanding and rotating flow $u^{\mu}$ defined in the Sec.~\eqref{fireball_rot} enters the local equilibrium distribution and modifies the particle spectra. The angular velocity parameter $\Omega(t)$ should decrease with time as the volume, or equivalently the momentum of inertia of the system, increases (see the expression of the net OAM) to keep the net OAM fixed along the $y$-axis. Therefore, even if the system starts with a higher angular velocity, its magnitude is diminished at the kinetic freeze-out time, i.e., $\Omega(t_{f})\equiv \Omega_{f} <\Omega(t=0)$. This is indeed the case for the more realistic transport simulations performed in Ref.~\cite{Jiang:2016woz}.
	\begin{figure*}[t]
			\centering
			\includegraphics[width=\linewidth]{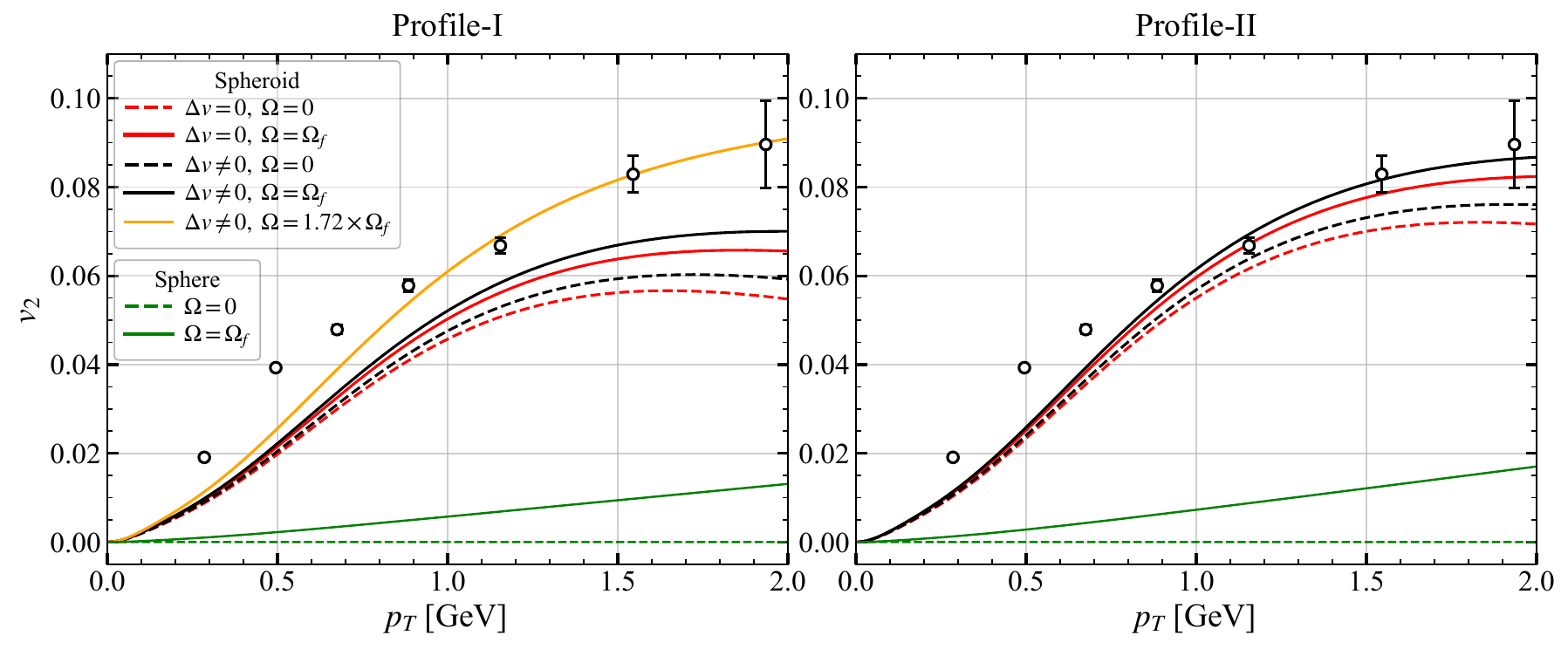}
		\caption{(Color online) $v_{2}$ for pion in the spheroid model for four different cases compared with the two cases of the spherical model for profile-I (left) and for profile-II ( right). $v_{2}$ are compared with STAR data~\cite{STAR:2013ayu} (scattered dots) for centrality $(0-80)\%$ at $\sqrt{s_{\rm NN}}=7.7$GeV. Both plots share the same legend.}
		\label{fig:pions_v2}
	\end{figure*}

As we have discussed in Sec.~\eqref{offcentral-OAM}, this rotational part of the $u^{\mu}$ has two observable effects which should be quantified in a detailed numerical analysis. Namely, the $p_{T}$-spectra at the higher end should get an increment, and the elliptic flow should be magnified due to the effects of rotation. For the time being, let us analyze it in a simple toy model by considering a globally rotating fluid~\cite{Becattini:2007nd}.  A globally rotating fluid has a distribution function given by,
\begin{eqnarray}
    f_{\Omega}&=&\frac{1}{e^{(E-\vec{u}\cdot\vec{p})/T_{0}} e^{-\mu_{0}/T_{0}}\pm 1}\nonumber\\
    &=&\frac{1}{e^{(E-(\vec{\Omega}\times \vec{r})\cdot\vec{p})/T_{0}} e^{-\mu_{0}/T_{0}}\pm 1}~,\label{rot_distr}
\end{eqnarray}
where we ignore the spin-orbit coupling, as we are not concerned with the effect of spin polarization in the present paper. Since the complete spectra given in Eq.~\eqref{eq:cooper_frye1} are proportional to the distribution function, the qualitative features of the spectra can be ascertained by analyzing the particle distribution function.
Notice that the distribution in Eq.~\eqref{rot_distr} contains a centrifuge effect, i.e., particles with their momenta in the $XZ$-plane would get an extra push from the rotation compared to the particles traveling in $y$-direction. 
\begin{figure*}[t]
		\centering
		\begin{subfigure}{0.48\textwidth}
			\centering
			\includegraphics[width=\linewidth]{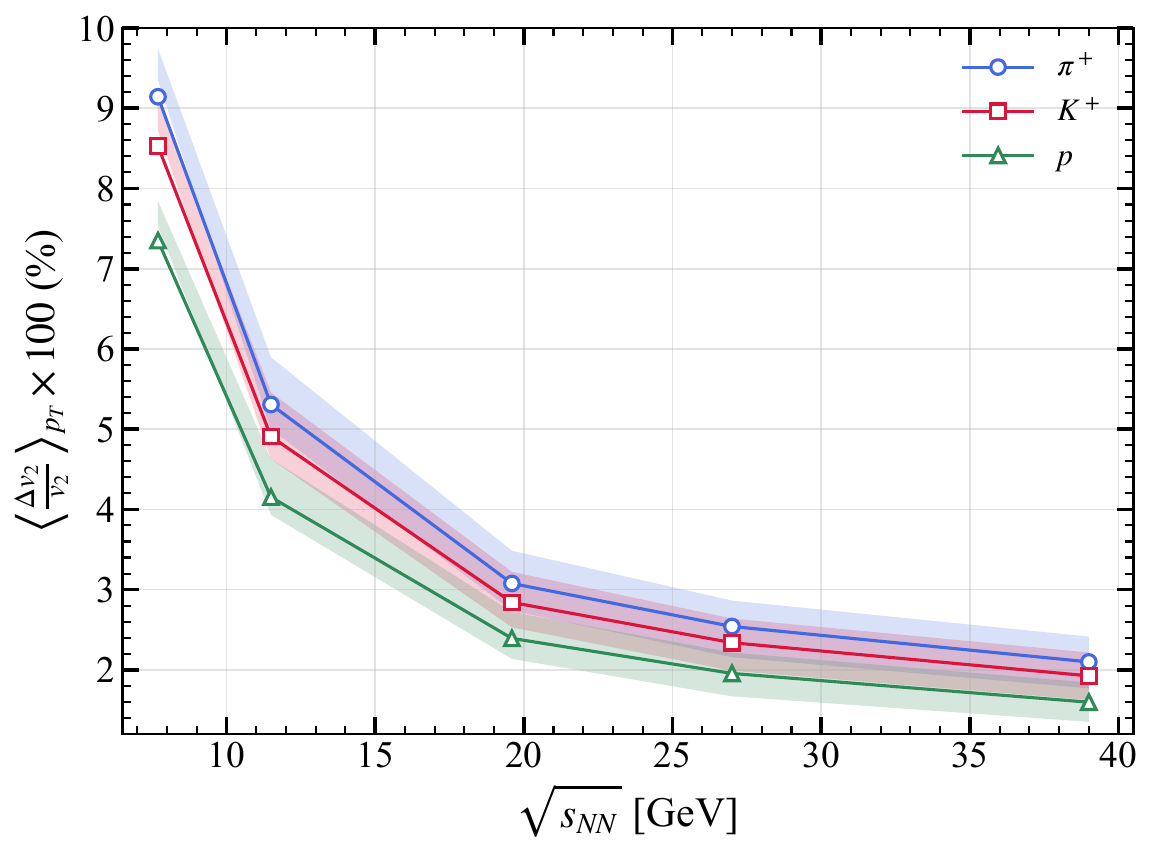}
			\label{fig:proton_pT}
		\end{subfigure}
		\begin{subfigure}{0.48\textwidth}
			\centering
			\includegraphics[width=\linewidth]{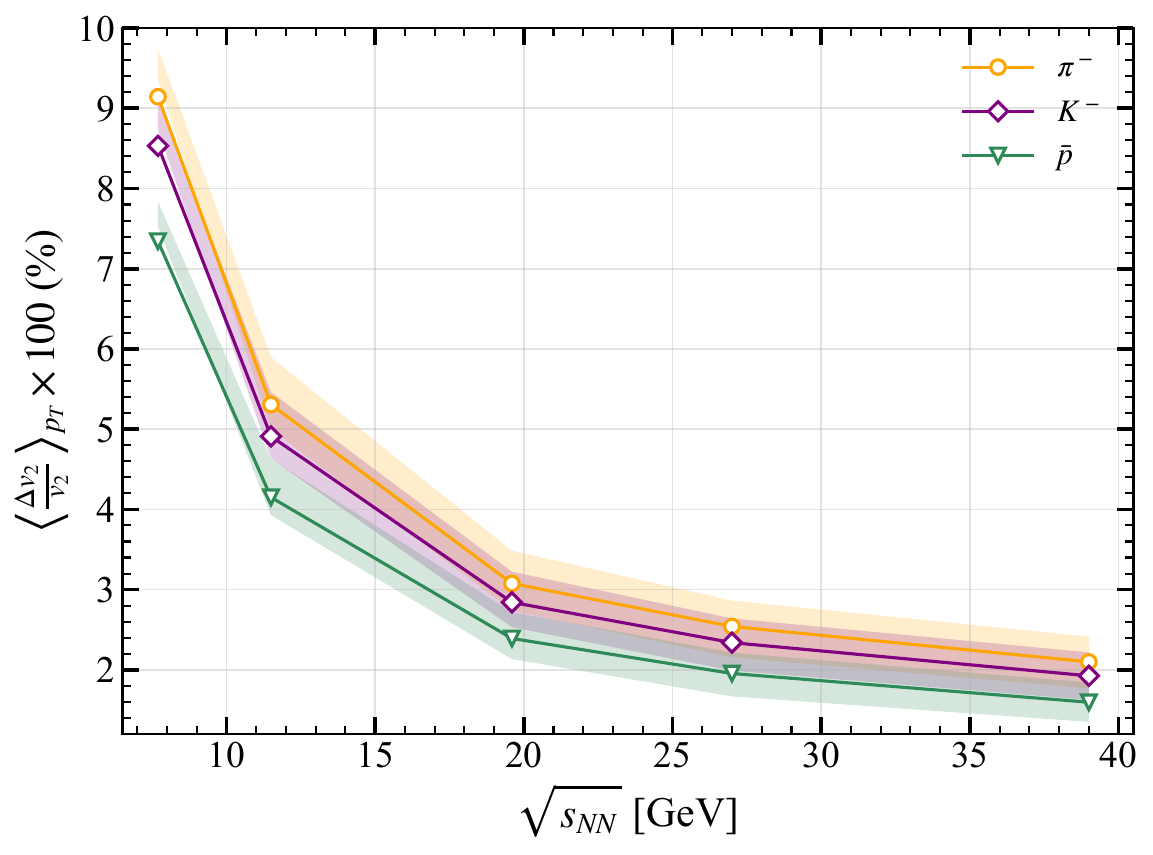}
			\label{fig:anti_proton_pT}
		\end{subfigure}
		\caption{(Color online) Calculated $\%$ change in $v_{2}$ averaged over the transverse momentum bin $0<p_{T}<2$ GeV for the identified hadrons in the RHIC BES $\sqrt{s_{\rm NN}}=7.7-39$ GeV for (a) particles (left) and (b) anti-particles (right) for mid central collisions.}
		\label{fig:all_v2_av_per}
	\end{figure*}
To exactly quantify the effect of rotation, we made the numerical analysis within the ambit of the models discussed in Sec.~\eqref{fireball_rot}. We use the value of angular velocity at kinetic freeze-out from the realistic AMPT (A Multi-Phase Transport) simulations~\cite{Jiang:2016woz}\footnote{We extrapolate their predictions to the low $\sqrt{s_{\rm NN}}$ using Eq.(8) of their paper. Also we keep a time offset of $1$ fm in Eq.(8) for the matter to equilibrate.}, in which the authors calculated the global angular velocity as a function of time by averaging the local vorticity over the system volume. Using the $\sqrt{s_{\rm NN}}$ and centrality dependent kinetic freeze-out time $t_{f}$ and initial system size which is motivated from our previous analysis~\cite{Rai:2026dda,Dwibedi:2026yhl}, we determine the model parameters $A$, $B$, $\Delta v$, and $v_{\infty}$. For each model, two different profiles described in Sec.~\eqref{fireball_rot} are considered. Specifically, we obtain the best-fit parameter values by matching the experimental data for the model spheroid-II, and then use those parameter sets unchanged in profile-I for both the sphere and the spheroid. This choice is made as the average angular velocity provided in Ref.~\cite{Jiang:2016woz} is a global average and independent of the $XY$-plane. The model parameters used are specified in Table~\eqref{table_kin}. In the spheroid model, there are two sources of momentum anisotropy induced by the velocity field: one is generated by the set $(\Delta v,B)$ together with the geometrical specifications, and the other is due to rotation. Setting $\Delta v=B=0$ and/or $\Omega=0$ would not turn off the momentum anisotropy because the inherent geometry still makes the velocity field anisotropic, i.e.,  $\Delta v=B=0$ makes the major and minor axis expansion rate the same, but they are still unequal because their initial lengths are different. In the spherical geometry, however, setting $\Omega=0$ should make the elliptic flow vanish as there are no additional sources of anisotropies. Therefore, one has four cases corresponding to $\Omega$ and $\Delta v$ zero and/or non-zero for the spheroid-I and II, and two cases for zero or non-zero values of $\Omega$ for sphere-I and II. We present these cases in relation to $p_{T}$-spectra and $v_{2}$, respectively, in Figs.~\eqref{fig:pion_pT} and \eqref{fig:pions_v2} for pion.
\begin{table}[H]
\centering
\caption{Parameter sets used for different collision energies. The parameters $A$, $B$, and $\Omega$ are in fm$^{-1}$, $b_0$ and $t_f$ are in fm, and $T$ is in MeV.
}
\setlength{\tabcolsep}{2pt}

\begin{tabular}{ccccccccc}
\hline\hline
$\sqrt{s_{\rm NN}}$ 
& $T$
& $A$
& $B$
& $v_{\infty}$
& $\Delta v$
& $b_0$
& $t_f$ 
& $\Omega_{f}$ 
\\
\hline
7.7  & 122 & 0.350 & 0.001 & 0.50 & 0.1 & 2.781 & 8.9 & 0.0112 \\
11.5 & 124 & 0.380 & 0.003 & 0.53 & 0.1 & 2.279 & 9.4 & 0.0098\\
19.6 & 129 & 0.430 & 0.005 & 0.54 & 0.1 & 1.745 & 10.0 & 0.0083\\
27   & 130 & 0.465 & 0.006 & 0.55 & 0.1 & 1.490 & 10.4 & 0.0077\\
39   & 133 & 0.489 & 0.007 & 0.56 & 0.1 & 1.230 & 10.5 & 0.0074\\
\hline\hline
\end{tabular}
\label{table_kin}
\end{table}

In Fig.~\eqref{fig:pion_pT}, we analyze the $p_{T}$ spectra, and there are 7 sets of curves in each plot, where 2 sets--1. $\Omega=0$ (green dashed line), 2. $\Omega=\Omega_{f}$ (green solid line) are for the spherical profile and 4 sets--3. $\Delta v=0, \Omega=0$ (red dashed line), 4. $\Delta v=0, \Omega=\Omega_{f}$ (red solid line), 5. $\Delta v\neq0, \Omega=0$ (black dashed line), 6. $\Delta v\neq0, \Omega=\Omega_{f}$ (black solid line) and 7. $\Delta v\neq0, \Omega=5\times \Omega_{f}$ (blue line) are for spheroidal profile.
The $p_{T}$ spectra show little sensitivity to the different profiles in the spheroidal model. This can be attributed to the smallness of $\Delta v$ as well as the $\Omega$ at kinetic freeze-out. Therefore, 4 sets of spheroid curves are almost aligned, and 2 sets of spheres are also not very separate. However, one can zoom in to see the difference by magnifying the value of $\Omega$.
By taking the $\Omega$ five times more than its expected value in the kinetic freeze-out, we can clearly see (blue lines in Fig.~\eqref{fig:pion_pT}) the effect where a large number of particles are obtained at higher $p_{T}$. Moreover, as we use the fit parameters of spheroid-II in the case of a sphere, this results in an enhanced spectrum at high $p_{T}$, which in turn implies that to explain the same $p_{T}$ spectra results in a spherical model, one needs a lower radial velocity, i.e., $\dot{r}_{B}(t_{f})$. 

However, we see that once the spectra are decomposed into Fourier components to evaluate $v_{2}$, all of these small changes significantly affect it and are clearly manifested in Fig.~\eqref{fig:pions_v2}. We see that all the momentum anisotropies contribute appreciably to $v_{2}$, with the ranking $v_{2}(\Delta v=\Omega=0)<v_{2}(\Delta v\neq0,\Omega=0)<v_{2}(\Delta v=0,\Omega\neq 0)<v_{2}(\Delta v\neq 0, \Omega \neq 0)$ for the spheroidal geometry. Evaluated $v_{2}(p_{T})$ for each different case in spheroid-I remains lower than spheroid-II. This is a manifestation of boost scaling in the $XZ$-plane velocities in spheroid-I compared with spheroid-II. Namely, since the $XZ$-plane velocities are scaled by the boost factor $\gamma_{y}=\cosh \eta_{y}$ along the $y$-axis, the same set of parameters predicts more $v_{2}$ in profile-II than in profile-I. Since angular velocity $\Omega_{I}=\Omega/\cosh\eta_{y}\leq \Omega$ one needs more $\Omega$ to generate the same $v_{2}$ as that of profile-II. This is shown by taking a 1.72 times larger value of $\Omega$ in the upper left panel of Fig.~\eqref{fig:pions_v2} in the orange line. The elliptic flow in the spherical geometry becomes zero in the absence of rotation, as there are no additional anisotropic mechanisms. Although we get a finite elliptic flow in the presence of $\Omega$, its magnitude lies significantly below the experimental data for the realistic magnitude of $\Omega_{f}$.

At last, we present what we believe are the main results, at least from the phenomenological point of view. Fig.~\eqref{fig:all_v2_av_per} displays the $p_{T}$-averaged change due to $\Omega_{f}$ in the $v_{2}$ for all identified hadrons in the RHIC BES regime $\sqrt{s_{\rm NN}}=7.7-39$ GeV. This is determined with the spheroid-II profile, which uses the space-averaged angular velocity from the AMPT model~\cite{Jiang:2016woz} (also provided in Table~\eqref{table_kin}). The elliptic flow for particles other than pions has been determined by first varying the particle-specific chemical potentials~\cite{Rai:2026dda,Dwibedi:2026yhl} to fit their spectra and using the other model parameters the same as those of pions, which corresponds to the same freeze-out environment for all hadrons. But the value particle specific chemical potential hardly matters for $v_{2}$ as at high temperature, Bose-Einstein as well as Fermi-Dirac distribution reduces to the Boltzmann distribution, so that the effect of $\mu$ cancels. Therefore, the $v_{2}$ is almost determined by the mass of the various hadrons. We see that the percentage change in the average $v_{2}$ for the particles and their corresponding antiparticles is nearly the same across beam energies. We observe a maximum change of about $9\%$ for pion at the beam energy $7.7$ GeV, which reduces to $2\%$ for $\sqrt{s_{\rm NN}}=39$ GeV. The percentage change for more massive particles (kaons and protons) is smaller than that of pions, as the $v_{2}(\Omega=0)$ follows the mass ordering with more massive particles leading to higher elliptic flow, whereas the relative change by incorporating angular velocity remains almost the same. As the angular velocity decreases with time, a lesser freeze-out time would lead to a higher angular velocity, thereby higher $v_{2}$. We check the sensitivity of the freeze-out time dependence by varying the freeze-out time $t_{f} \pm 1.5$ fm. The percentage in $v_{2}$ is observed in this way within $\pm 1\%$.

Overall, our results demonstrate that azimuthal anisotropy in the momentum spectra, and hence elliptic flow, can arise from two distinct sources: anisotropic expansion in the transverse plane and rotation about the $y$-axis. For a spherical fireball model, the expanding fluid flow, derived from the rate of increase of the fireball radius, is isotropic. The elliptic flow in such scenarios originates entirely from the momentum anisotropy induced by the rotating flow. Although a sufficiently large angular velocity could, in principle, generate elliptic flow comparable to the experimentally observed magnitude, the angular velocity at freeze-out is constrained by the conservation of OAM. For the realistic values, $\Omega\sim10^{-2}$~fm$^{-1}$, the rotational contribution therefore amounts to only a few percent of the observed elliptic flow. In contrast, the spheroidal geometry incorporates anisotropic transverse expansion, together with the rotational flow, which provides two complementary sources of elliptic flow and is consistent with the experimental data. Thus, within our fireball framework, anisotropic expansion provides the dominant contribution to elliptic flow, while vorticity associated with the initial OAM provides a non-negligible additional contribution.

\section{SUMMARY AND OUTLOOK}\label{summary}
In this work, we have investigated how the orbital angular momentum carried by matter produced in off-central heavy-ion collisions can leave an imprint on the final-state anisotropic momentum spectra. We generalized the traditional blast-wave framework by incorporating a rotational component into the expanding fluid velocity. This results in the construction of fireball profiles with non-zero orbital angular momentum. Two forms of rotation were considered: global rotation and differential rotation, in which different layers of the fireball rotate with different angular velocities. Using spherical and spheroidal fireball geometries, we disentangled the momentum anisotropy generated by rotation and anisotropic expansion. Our results show that, for realistic angular velocities constrained by the orbital angular momentum carried by the system, rotation provides a non-negligible contribution to elliptic flow, while anisotropic expansion remains the dominant source. In the spheroidal geometry, the interplay of these two mechanisms leads to a $p_T$-dependent enhancement of $v_2(p_T)$ by about $7-15\%$ at $\sqrt{s_{\rm NN}}=7.7$ GeV due to rotation, corresponding to an average enhancement of approximately $7-9\%$ over the $p_T=0-2$ GeV range.
 The angular velocities used in our calculation are obtained from transport simulations and are larger at lower collision energies, indicating that the rotation-induced contribution can become relatively more important toward lower beam energies.

The present study demonstrates for the first time that a rotating blast-wave framework, motivated by the orbital angular momentum carried by the system, provides a simple and physically motivated approach to phenomenological descriptions of off-central heavy-ion collisions. More studies in the direction of rotation-sensitive observables could provide further constraints on the rotational dynamics of the fireball and help quantify the role of vorticity in the evolution of strongly interacting matter. 

\section{ACKNOWLEDGEMENT}
This work was supported in part by the Board of Research in Nuclear Sciences (BRNS) and the Department of Atomic Energy (DAE), Government of India, with Grant Nos. 57/14/01/2024-BRNS/313 (S.G.) and the Ministry of Education, Government of India (A.D.). 
\appendix
\bibliographystyle{unsrturl}
\bibliography{ref}

@article{STAR:2013ayu,
    author = "Adamczyk, L. and others",
    collaboration = "STAR",
    title = "{Elliptic flow of identified hadrons in Au+Au collisions at $\sqrt{s_{\rm NN}}=$ 7.7-62.4 GeV}",
    eprint = "1301.2348",
    archivePrefix = "arXiv",
    primaryClass = "nucl-ex",
    doi = "10.1103/PhysRevC.88.014902",
    journal = "Phys. Rev. C",
    volume = "88",
    pages = "014902",
    year = "2013"
}

@article{Gale:2013da,
	author = "Gale, Charles and Jeon, Sangyong and Schenke, Bjoern",
	title = "{Hydrodynamic Modeling of Heavy-Ion Collisions}",
	eprint = "1301.5893",
	archivePrefix = "arXiv",
	primaryClass = "nucl-th",
	doi = "10.1142/S0217751X13400113",
	journal = "Int. J. Mod. Phys. A",
	volume = "28",
	pages = "1340011",
	year = "2013"
}

@article{De:2022yxq,
	author = "De, A. and Kapusta, J. I. and Singh, M. and Welle, T.",
	title = "{Comprehensive simulation of heavy-ion collisions at nonzero baryon chemical potential}",
	eprint = "2206.02655",
	archivePrefix = "arXiv",
	primaryClass = "nucl-th",
	doi = "10.1103/PhysRevC.106.054906",
	journal = "Phys. Rev. C",
	volume = "106",
	number = "5",
	pages = "054906",
	year = "2022"
}

@book{Romatschke:2017ejr,
	author = "Romatschke, Paul and Romatschke, Ulrike",
	title = "{Relativistic Fluid Dynamics In and Out of Equilibrium}",
	eprint = "1712.05815",
	archivePrefix = "arXiv",
	primaryClass = "nucl-th",
	doi = "10.1017/9781108651998",
	isbn = "978-1-108-48368-1, 978-1-108-75002-8",
	publisher = "Cambridge University Press",
	series = "Cambridge Monographs on Mathematical Physics",
	month = "5",
	year = "2019"
}

@article{10.1093/ptep/pts014,
	author = {Nonaka, Chiho and Asakawa, Masayuki},
	title = {Modeling a realistic dynamical model for high energy heavy ion collisions},
	journal = {Progress of Theoretical and Experimental Physics},
	volume = {2012},
	number = {1},
	pages = {01A208},
	year = {2012},
	month = {09},
	issn = {2050-3911},
	doi = {10.1093/ptep/pts014},
	url = {https://doi.org/10.1093/ptep/pts014}
}

@article{Huovinen:2001cy,
    author = "Huovinen, P. and Kolb, P. F. and Heinz, Ulrich W. and Ruuskanen, P. V. and Voloshin, S. A.",
    title = "{Radial and elliptic flow at RHIC: Further predictions}",
    eprint = "hep-ph/0101136",
    archivePrefix = "arXiv",
    doi = "10.1016/S0370-2693(01)00219-2",
    journal = "Phys. Lett. B",
    volume = "503",
    pages = "58--64",
    year = "2001"
}

@article{Heinz:2013th,
    author = "Heinz, Ulrich and Snellings, Raimond",
    title = "{Collective flow and viscosity in relativistic heavy-ion collisions}",
    eprint = "1301.2826",
    archivePrefix = "arXiv",
    primaryClass = "nucl-th",
    doi = "10.1146/annurev-nucl-102212-170540",
    journal = "Ann. Rev. Nucl. Part. Sci.",
    volume = "63",
    pages = "123--151",
    year = "2013"
}

@article{Harris:2023tti,
	author = {Harris, John W. and M{\"u}ller, Berndt},
	title = "{''QGP Signatures'' Revisited}",
	eprint = "2308.05743",
	archivePrefix = "arXiv",
	primaryClass = "hep-ph",
	doi = "10.1140/epjc/s10052-024-12533-y",
	journal = "Eur. Phys. J. C",
	volume = "84",
	number = "3",
	pages = "247",
	year = "2024"
}

@article{Kolb:2003dz,
	author = "Kolb, Peter F. and Heinz, Ulrich W.",
	editor = "Hwa, Rudolph C. and Wang, Xin-Nian",
	title = "{Hydrodynamic description of ultrarelativistic heavy ion collisions}",
	eprint = "nucl-th/0305084",
	archivePrefix = "arXiv",
	reportNumber = "SUNY-NTG-03-06",
	pages = "634--714",
	month = "5",
	year = "2003"
}

@article{STAR:2017sal,
	author = "Adamczyk, L. and others",
	collaboration = "STAR",
	title = "{Bulk Properties of the Medium Produced in Relativistic Heavy-Ion Collisions from the Beam Energy Scan Program}",
	eprint = "1701.07065",
	archivePrefix = "arXiv",
	primaryClass = "nucl-ex",
	doi = "10.1103/PhysRevC.96.044904",
	journal = "Phys. Rev. C",
	volume = "96",
	number = "4",
	pages = "044904",
	year = "2017"
}

@article{Cooper:1974mv,
	author = "Cooper, Fred and Frye, Graham",
	title = "{Comment on the Single Particle Distribution in the Hydrodynamic and Statistical Thermodynamic Models of Multiparticle Production}",
	reportNumber = "Print-74-0742 (YESHIVA)",
	doi = "10.1103/PhysRevD.10.186",
	journal = "Phys. Rev. D",
	volume = "10",
	pages = "186",
	year = "1974"
}

@article{Ali:2024zvp,
	author = "Ali, Mahammad Sabir and Biswas, Deeptak and Jaiswal, Amaresh and Singh, Sushant K.",
	title = "{Hadron momentum spectra from analytical solutions of relativistic hydrodynamics}",
	eprint = "2403.00624",
	archivePrefix = "arXiv",
	primaryClass = "hep-ph",
	doi = "10.1140/epjc/s10052-025-13751-8",
	journal = "Eur. Phys. J. C",
	volume = "85",
	number = "1",
	pages = "30",
	year = "2025"
}

@article{Florkowski:2004tn,
	author = "Florkowski, Wojciech and Broniowski, Wojciech",
	editor = "Sadzikowski, M.",
	title = "{Hydro-inspired parameterizations of freeze-out in relativistic heavy-ion collisions}",
	eprint = "nucl-th/0410081",
	archivePrefix = "arXiv",
	journal = "Acta Phys. Polon. B",
	volume = "35",
	pages = "2895--2910",
	year = "2004"
}

@article{Schnedermann:1993ws,
	author = "Schnedermann, Ekkard and Sollfrank, Josef and Heinz, Ulrich W.",
	title = "{Thermal phenomenology of hadrons from 200-A/GeV S+S collisions}",
	eprint = "nucl-th/9307020",
	archivePrefix = "arXiv",
	reportNumber = "TPR-93-16",
	doi = "10.1103/PhysRevC.48.2462",
	journal = "Phys. Rev. C",
	volume = "48",
	pages = "2462--2475",
	year = "1993"
}

@article{Retiere:2003kf,
	author = "Retiere, Fabrice and Lisa, Michael Annan",
	title = "{Observable implications of geometrical and dynamical aspects of freeze out in heavy ion collisions}",
	eprint = "nucl-th/0312024",
	archivePrefix = "arXiv",
	doi = "10.1103/PhysRevC.70.044907",
	journal = "Phys. Rev. C",
	volume = "70",
	pages = "044907",
	year = "2004"
}

@article{Tomasik:2024uuq,
	author = "Tomasik, Boris",
	title = "{On elliptic flow and the blast-wave model}",
	eprint = "2409.19758",
	archivePrefix = "arXiv",
	primaryClass = "nucl-th",
	doi = "10.1142/S0217751X25420060",
	journal = "Int. J. Mod. Phys. A",
	volume = "40",
	number = "21",
	pages = "2542006",
	year = "2025"
}

@article{Broniowski:2001we,
	author = "Broniowski, Wojciech and Florkowski, Wojciech",
	title = "{Explanation of the RHIC p(T) spectra in a thermal model with expansion}",
	eprint = "nucl-th/0106050",
	archivePrefix = "arXiv",
	doi = "10.1103/PhysRevLett.87.272302",
	journal = "Phys. Rev. Lett.",
	volume = "87",
	pages = "272302",
	year = "2001"
}

@article{PhysRevLett.42.880,
	title = {Evidence for a Blast Wave from Compressed Nuclear Matter},
	author = {Siemens, Philip J. and Rasmussen, John O.},
	journal = {Phys. Rev. Lett.},
	volume = {42},
	issue = {14},
	pages = {880--883},
	numpages = {0},
	year = {1979},
	month = {Apr},
	publisher = {American Physical Society},
	doi = {10.1103/PhysRevLett.42.880},
	url = {https://link.aps.org/doi/10.1103/PhysRevLett.42.880}
}

@article{Harabasz:2020sei,
	author = "Harabasz, Szymon and Florkowski, Wojciech and Galatyuk, Tetyana and Ma Lgorzata Gumberidze, {\textdaggerdbl}. and Ryblewski, Radoslaw and Salabura, Piotr and Stroth, Joachim",
	title = "{Statistical hadronization model for heavy-ion collisions in the few-GeV energy regime}",
	eprint = "2003.12992",
	archivePrefix = "arXiv",
	primaryClass = "nucl-th",
	doi = "10.1103/PhysRevC.102.054903",
	journal = "Phys. Rev. C",
	volume = "102",
	number = "5",
	pages = "054903",
	year = "2020"
}

@article{Harabasz:2022rdt,
	author = "Harabasz, Szymon and Ko{\l}a{\'s}, Jedrzej and Ryblewski, Rados{\l}aw and Florkowski, Wojciech and Galatyuk, Tetyana and Gumberidze, Ma{\l}gorzata and Salabura, Piotr and Stroth, Joachim and Zbroszczyk, Hanna Paulina",
	title = "{Spheroidal expansion and freeze-out geometry of heavy-ion collisions in the few-GeV energy regime}",
	eprint = "2210.07694",
	archivePrefix = "arXiv",
	primaryClass = "nucl-th",
	doi = "10.1103/PhysRevC.107.034917",
	journal = "Phys. Rev. C",
	volume = "107",
	number = "3",
	pages = "034917",
	year = "2023"
}

@article{Broniowski:2001uk,
	author = "Broniowski, Wojciech and Florkowski, Wojciech",
	title = "{Strange particle production at RHIC in a single freezeout model}",
	eprint = "nucl-th/0112043",
	archivePrefix = "arXiv",
	doi = "10.1103/PhysRevC.65.064905",
	journal = "Phys. Rev. C",
	volume = "65",
	pages = "064905",
	year = "2002"
}

@article{Alam:2026ixb,
	author = "Alam, Sk Noor and Roy, Victor",
	title = "{Kinetic Freeze-Out Conditions and Net Baryon Density in Au+Au Collisions at $\sqrt{s_{NN}} = 7.7$--$39$ GeV within a Collective Flow Fireball Model}",
	eprint = "2603.07160",
	archivePrefix = "arXiv",
	primaryClass = "nucl-th",
	month = "3",
	year = "2026"
}

@article{Rai:2026dda,
	author = "Rai, Anand and Dwibedi, Ashutosh and Ghosh, Sabyasachi",
	title = "{Spectra and elliptic flow of light hadrons in an expanding fire-cylinder model for the RHIC Beam Energy Scan}",
	eprint = "2602.17241",
	archivePrefix = "arXiv",
	primaryClass = "nucl-th",
	doi = "10.1016/j.nuclphysa.2026.123440",
	journal = "Nucl. Phys. A",
	volume = "1073",
	pages = "123440",
	year = "2026"
}

@article{Dwibedi:2026yhl,
    author = "Dwibedi, Ashutosh and Panja, Anupam and Ghosh, Sabyasachi",
    title = "{An expanding spherical fireball model for light hadron production at RHIC ($\sqrt{s_{\rm NN}}=7.7$--$39$ GeV)}",
    eprint = "2607.04191",
    archivePrefix = "arXiv",
    primaryClass = "nucl-th",
    month = "7",
    year = "2026"
}

@article{Florkowski:2019voj,
    author = "Florkowski, Wojciech and Kumar, Avdhesh and Ryblewski, Radoslaw and Mazeliauskas, Aleksas",
    title = "{Longitudinal spin polarization in a thermal model}",
    eprint = "1904.00002",
    archivePrefix = "arXiv",
    primaryClass = "nucl-th",
    doi = "10.1103/PhysRevC.100.054907",
    journal = "Phys. Rev. C",
    volume = "100",
    number = "5",
    pages = "054907",
    year = "2019"
}

@article{Florkowski:2021xvy,
    author = "Florkowski, Wojciech and Kumar, Avdhesh and Mazeliauskas, Aleksas and Ryblewski, Radoslaw",
    title = "{Effect of thermal shear on longitudinal spin polarization in a thermal model}",
    eprint = "2112.02799",
    archivePrefix = "arXiv",
    primaryClass = "hep-ph",
    doi = "10.1103/PhysRevC.105.064901",
    journal = "Phys. Rev. C",
    volume = "105",
    number = "6",
    pages = "064901",
    year = "2022"
}

@article{Banerjee:2024xnd,
    author = "Banerjee, Soham and Bhadury, Samapan and Florkowski, Wojciech and Jaiswal, Amaresh and Ryblewski, Radoslaw",
    title = "{Longitudinal spin polarization in a thermal model with dissipative corrections}",
    eprint = "2405.05089",
    archivePrefix = "arXiv",
    primaryClass = "hep-ph",
    doi = "10.1103/923l-yxkc",
    journal = "Phys. Rev. C",
    volume = "111",
    number = "6",
    pages = "064912",
    year = "2025"
}

@article{Gossiaux:2011ea,
	author = "Gossiaux, Pol Bernard and Vogel, Sascha and van Hees, Hendrik and Aichelin, Joerg and Rapp, Ralf and He, Min and Bluhm, Marcus",
	title = "{The Influence of bulk evolution models on heavy-quark phenomenology}",
	eprint = "1102.1114",
	archivePrefix = "arXiv",
	primaryClass = "hep-ph",
	month = "2",
	year = "2011"
}

@article{Bass:1998ca,
	author = "Bass, S. A. and others",
	title = "{Microscopic models for ultrarelativistic heavy ion collisions}",
	eprint = "nucl-th/9803035",
	archivePrefix = "arXiv",
	doi = "10.1016/S0146-6410(98)00058-1",
	journal = "Prog. Part. Nucl. Phys.",
	volume = "41",
	pages = "255--369",
	year = "1998"
}

@article{Bondorf:1978kz,
	author = "Bondorf, J. P. and Garpman, S. I. A. and Zimanyi, J.",
	title = "{A Simple Analytic Hydrodynamic Model for Expanding Fireballs}",
	doi = "10.1016/0375-9474(78)90076-3",
	journal = "Nucl. Phys. A",
	volume = "296",
	pages = "320--332",
	year = "1978"
}

@article{Liang:2004ph,
    author = "Liang, Zuo-Tang and Wang, Xin-Nian",
    title = "{Globally polarized quark-gluon plasma in non-central A+A collisions}",
    eprint = "nucl-th/0410079",
    archivePrefix = "arXiv",
    reportNumber = "LBNL-56383",
    doi = "10.1103/PhysRevLett.94.102301",
    journal = "Phys. Rev. Lett.",
    volume = "94",
    pages = "102301",
    year = "2005",
    note = "[Erratum: Phys.Rev.Lett. 96, 039901 (2006)]"
}

@article{Gao:2007bc,
    author = "Gao, Jian-Hua and Chen, Shou-Wan and Deng, Wei-tian and Liang, Zuo-Tang and Wang, Qun and Wang, Xin-Nian",
    title = "{Global quark polarization in non-central A+A collisions}",
    eprint = "0710.2943",
    archivePrefix = "arXiv",
    primaryClass = "nucl-th",
    reportNumber = "LBNL-63515",
    doi = "10.1103/PhysRevC.77.044902",
    journal = "Phys. Rev. C",
    volume = "77",
    pages = "044902",
    year = "2008"
}

@article{Betz:2007kg,
    author = "Betz, Barbara and Gyulassy, Miklos and Torrieri, Giorgio",
    title = "{Polarization probes of vorticity in heavy ion collisions}",
    eprint = "0708.0035",
    archivePrefix = "arXiv",
    primaryClass = "nucl-th",
    doi = "10.1103/PhysRevC.76.044901",
    journal = "Phys. Rev. C",
    volume = "76",
    pages = "044901",
    year = "2007"
}

@article{Becattini:2007sr,
    author = "Becattini, F. and Piccinini, F. and Rizzo, J.",
    title = "{Angular momentum conservation in heavy ion collisions at very high energy}",
    eprint = "0711.1253",
    archivePrefix = "arXiv",
    primaryClass = "nucl-th",
    doi = "10.1103/PhysRevC.77.024906",
    journal = "Phys. Rev. C",
    volume = "77",
    pages = "024906",
    year = "2008"
}

@article{Becattini:2007nd,
    author = "Becattini, F. and Piccinini, F.",
    title = "{The Ideal relativistic spinning gas: Polarization and spectra}",
    eprint = "0710.5694",
    archivePrefix = "arXiv",
    primaryClass = "nucl-th",
    doi = "10.1016/j.aop.2008.01.001",
    journal = "Annals Phys.",
    volume = "323",
    pages = "2452--2473",
    year = "2008"
}

@article{Becattini:2013fla,
    author = "Becattini, F. and Chandra, V. and Del Zanna, L. and Grossi, E.",
    title = "{Relativistic distribution function for particles with spin at local thermodynamical equilibrium}",
    eprint = "1303.3431",
    archivePrefix = "arXiv",
    primaryClass = "nucl-th",
    doi = "10.1016/j.aop.2013.07.004",
    journal = "Annals Phys.",
    volume = "338",
    pages = "32--49",
    year = "2013"
}

@article{Becattini:2013vja,
    author = "Becattini, F. and Csernai, L. and Wang, D. J.",
    title = "{$\Lambda$ polarization in peripheral heavy ion collisions}",
    eprint = "1304.4427",
    archivePrefix = "arXiv",
    primaryClass = "nucl-th",
    doi = "10.1103/PhysRevC.88.034905",
    journal = "Phys. Rev. C",
    volume = "88",
    number = "3",
    pages = "034905",
    year = "2013",
    note = "[Erratum: Phys.Rev.C 93, 069901 (2016)]"
}

@article{Csernai:2013bqa,
    author = "Csernai, L. P. and Magas, V. K. and Wang, D. J.",
    title = "{Flow Vorticity in Peripheral High Energy Heavy Ion Collisions}",
    eprint = "1302.5310",
    archivePrefix = "arXiv",
    primaryClass = "nucl-th",
    doi = "10.1103/PhysRevC.87.034906",
    journal = "Phys. Rev. C",
    volume = "87",
    number = "3",
    pages = "034906",
    year = "2013"
}

@article{Becattini:2015ska,
    author = "Becattini, F. and Inghirami, G. and Rolando, V. and Beraudo, A. and Del Zanna, L. and De Pace, A. and Nardi, M. and Pagliara, G. and Chandra, V.",
    title = "{A study of vorticity formation in high energy nuclear collisions}",
    eprint = "1501.04468",
    archivePrefix = "arXiv",
    primaryClass = "nucl-th",
    doi = "10.1140/epjc/s10052-015-3624-1",
    journal = "Eur. Phys. J. C",
    volume = "75",
    number = "9",
    pages = "406",
    year = "2015",
    note = "[Erratum: Eur.Phys.J.C 78, 354 (2018)]"
}

@article{Karpenko:2016jyx,
    author = "Karpenko, I. and Becattini, F.",
    title = "{Study of $\Lambda $ polarization in relativistic nuclear collisions at $\sqrt{s_\mathrm {NN}}=7.7$ {\textendash}200 GeV}",
    eprint = "1610.04717",
    archivePrefix = "arXiv",
    primaryClass = "nucl-th",
    doi = "10.1140/epjc/s10052-017-4765-1",
    journal = "Eur. Phys. J. C",
    volume = "77",
    number = "4",
    pages = "213",
    year = "2017"
}

@inbook{Karpenko:2021wdm,
    author = "Karpenko, Iurii",
    title = "{Vorticity and Polarization in Heavy-Ion Collisions: Hydrodynamic Models}",
    eprint = "2101.04963",
    archivePrefix = "arXiv",
    primaryClass = "nucl-th",
    doi = "10.1007/978-3-030-71427-7_8",
    year = "2021"
}

@article{Deng:2016gyh,
    author = "Deng, Wei-Tian and Huang, Xu-Guang",
    title = "{Vorticity in Heavy-Ion Collisions}",
    eprint = "1603.06117",
    archivePrefix = "arXiv",
    primaryClass = "nucl-th",
    doi = "10.1103/PhysRevC.93.064907",
    journal = "Phys. Rev. C",
    volume = "93",
    number = "6",
    pages = "064907",
    year = "2016"
}

@article{Jiang:2016woz,
    author = "Jiang, Yin and Lin, Zi-Wei and Liao, Jinfeng",
    title = "{Rotating quark-gluon plasma in relativistic heavy ion collisions}",
    eprint = "1602.06580",
    archivePrefix = "arXiv",
    primaryClass = "hep-ph",
    doi = "10.1103/PhysRevC.94.044910",
    journal = "Phys. Rev. C",
    volume = "94",
    number = "4",
    pages = "044910",
    year = "2016",
    note = "[Erratum: Phys.Rev.C 95, 049904 (2017)]"
    }

@article{Huang:2020dtn,
    author = "Huang, Xu-Guang and Liao, Jinfeng and Wang, Qun and Xia, Xiao-Liang",
    title = "{Vorticity and Spin Polarization in Heavy Ion Collisions: Transport Models}",
    eprint = "2010.08937",
    archivePrefix = "arXiv",
    primaryClass = "nucl-th",
    doi = "10.1007/978-3-030-71427-7_9",
    journal = "Lect. Notes Phys.",
    volume = "987",
    pages = "281--308",
    year = "2021"
}

@article{Becattini:2020ngo,
    author = "Becattini, Francesco and Lisa, Michael A.",
    title = "{Polarization and Vorticity in the Quark{\textendash}Gluon Plasma}",
    eprint = "2003.03640",
    archivePrefix = "arXiv",
    primaryClass = "nucl-ex",
    doi = "10.1146/annurev-nucl-021920-095245",
    journal = "Ann. Rev. Nucl. Part. Sci.",
    volume = "70",
    pages = "395--423",
    year = "2020"
}

@book{Becattini:2021Strongly,
  editor    = {Becattini, F. and Liao, J. and Lisa, M. A.},
  title     = {Strongly Interacting Matter under Rotation},
  series    = {Lecture Notes in Physics},
  volume    = {987},
  publisher = {Springer},
  year      = {2021},
  doi       = {10.1007/978-3-030-71427-7}
}

@article{Niida:2024ntm,
    author = "Niida, Takafumi and Voloshin, Sergei A.",
    title = "{Polarization phenomenon in heavy-ion collisions}",
    eprint = "2404.11042",
    archivePrefix = "arXiv",
    primaryClass = "nucl-ex",
    doi = "10.1142/S0218301324300108",
    journal = "Int. J. Mod. Phys. E",
    volume = "33",
    number = "09",
    pages = "2430010",
    year = "2024"
}

@article{Becattini:2024uha,
    author = "Becattini, Francesco and Buzzegoli, Matteo and Niida, Takafumi and Pu, Shi and Tang, Ai-Hong and Wang, Qun",
    title = "{Spin polarization in relativistic heavy-ion collisions}",
    eprint = "2402.04540",
    archivePrefix = "arXiv",
    primaryClass = "nucl-th",
    doi = "10.1142/9789811294679_0005",
    journal = "Int. J. Mod. Phys. E",
    volume = "33",
    number = "06",
    pages = "2430006",
    year = "2024"
}

@article{Dey:2026epy,
    author = "Dey, Sourav and Das, Arpan and Mishra, Hiranmaya and Jaiswal, Amaresh",
    title = "{Spin dynamics and polarization in relativistic systems: recent developments}",
    eprint = "2605.12554",
    archivePrefix = "arXiv",
    primaryClass = "nucl-th",
    month = "5",
    year = "2026"
}

@article{STAR:2017ckg,
    author = "Adamczyk, L. and others",
    collaboration = "STAR",
    title = "{Global $\Lambda$ hyperon polarization in nuclear collisions: evidence for the most vortical fluid}",
    eprint = "1701.06657",
    archivePrefix = "arXiv",
    primaryClass = "nucl-ex",
    doi = "10.1038/nature23004",
    journal = "Nature",
    volume = "548",
    pages = "62--65",
    year = "2017"
}

@article{STAR:2018gyt,
    author = "Adam, Jaroslav and others",
    collaboration = "STAR",
    title = "{Global polarization of $\Lambda$ hyperons in Au+Au collisions at $\sqrt{s_{_{NN}}}$ = 200 GeV}",
    eprint = "1805.04400",
    archivePrefix = "arXiv",
    primaryClass = "nucl-ex",
    doi = "10.1103/PhysRevC.98.014910",
    journal = "Phys. Rev. C",
    volume = "98",
    pages = "014910",
    year = "2018"
}

@article{STAR:2019erd,
    author = "Adam, Jaroslav and others",
    collaboration = "STAR",
    title = "{Polarization of $\Lambda$ ($\bar{\Lambda}$) hyperons along the beam direction in Au+Au collisions at $\sqrt{s_{_{NN}}}$ = 200 GeV}",
    eprint = "1905.11917",
    archivePrefix = "arXiv",
    primaryClass = "nucl-ex",
    doi = "10.1103/PhysRevLett.123.132301",
    journal = "Phys. Rev. Lett.",
    volume = "123",
    number = "13",
    pages = "132301",
    year = "2019"
}

@article{STAR:2020xbm,
    author = "Adam, J. and others",
    collaboration = "STAR",
    title = "{Global Polarization of $\Xi$ and $\Omega$ Hyperons in Au+Au Collisions at $\sqrt {s_{NN}}$ = 200  GeV}",
    eprint = "2012.13601",
    archivePrefix = "arXiv",
    primaryClass = "nucl-ex",
    doi = "10.1103/PhysRevLett.126.162301",
    journal = "Phys. Rev. Lett.",
    volume = "126",
    number = "16",
    pages = "162301",
    year = "2021",
    note = "[Erratum: Phys.Rev.Lett. 131, 089901 (2023)]"
}

@article{STAR:2021beb,
    author = "Abdallah, M. S. and others",
    collaboration = "STAR",
    title = "{Global $\Lambda$-hyperon polarization in Au+Au collisions at $\sqrt {s_{NN}}$=3~GeV}",
    eprint = "2108.00044",
    archivePrefix = "arXiv",
    primaryClass = "nucl-ex",
    doi = "10.1103/PhysRevC.104.L061901",
    journal = "Phys. Rev. C",
    volume = "104",
    number = "6",
    pages = "L061901",
    year = "2021"
}

@article{STAR:2023eck,
    author = "Abdulhamid, Muhammad and others",
    collaboration = "STAR",
    title = "{Hyperon Polarization along the Beam Direction Relative to the Second and Third Harmonic Event Planes in Isobar Collisions at sNN=200{\,}{\,}GeV}",
    eprint = "2303.09074",
    archivePrefix = "arXiv",
    primaryClass = "nucl-ex",
    doi = "10.1103/PhysRevLett.131.202301",
    journal = "Phys. Rev. Lett.",
    volume = "131",
    number = "20",
    pages = "202301",
    year = "2023"
}

@article{STAR:2023nvo,
    author = "Abdulhamid, M. I. and others",
    collaboration = "STAR",
    title = "{Global polarization of {\ensuremath{\Lambda}} and {\ensuremath{\Lambda}}{\textasciimacron} hyperons in Au+Au collisions at sNN=19.6 and 27 GeV}",
    eprint = "2305.08705",
    archivePrefix = "arXiv",
    primaryClass = "nucl-ex",
    doi = "10.1103/PhysRevC.108.014910",
    journal = "Phys. Rev. C",
    volume = "108",
    number = "1",
    pages = "014910",
    year = "2023"
}

@article{ALICE:2021pzu,
    author = "Acharya, Shreyasi and others",
    collaboration = "ALICE",
    title = "{Polarization of $\Lambda$ and $\bar \Lambda$ Hyperons along the Beam Direction in Pb-Pb Collisions at $\sqrt {s_{NN}}$=5.02{\,}{\,}TeV}",
    eprint = "2107.11183",
    archivePrefix = "arXiv",
    primaryClass = "nucl-ex",
    reportNumber = "CERN-EP-2021-148",
    doi = "10.1103/PhysRevLett.128.172005",
    journal = "Phys. Rev. Lett.",
    volume = "128",
    number = "17",
    pages = "172005",
    year = "2022"
}

@article{ALICE:2022dyy,
    author = "Acharya, Shreyasi and others",
    collaboration = "ALICE",
    title = "{Measurement of the J/{\ensuremath{\psi}} Polarization with Respect to the Event Plane in Pb-Pb Collisions at the LHC}",
    eprint = "2204.10171",
    archivePrefix = "arXiv",
    primaryClass = "nucl-ex",
    reportNumber = "CERN-EP-2022-066",
    doi = "10.1103/PhysRevLett.131.042303",
    journal = "Phys. Rev. Lett.",
    volume = "131",
    number = "4",
    pages = "042303",
    year = "2023"
}

@article{STAR:2022fan,
    author = "Abdallah, M. S. and others",
    collaboration = "STAR",
    title = "{Pattern of global spin alignment of {\ensuremath{\phi}} and K$^{*0}$ mesons in heavy-ion collisions}",
    eprint = "2204.02302",
    archivePrefix = "arXiv",
    primaryClass = "hep-ph",
    doi = "10.1038/s41586-022-05557-5",
    journal = "Nature",
    volume = "614",
    number = "7947",
    pages = "244--248",
    year = "2023"
}

@article{Lin:2004en,
    author = "Lin, Zi-Wei and Ko, Che Ming and Li, Bao-An and Zhang, Bin and Pal, Subrata",
    title = "{A Multi-phase transport model for relativistic heavy ion collisions}",
    eprint = "nucl-th/0411110",
    archivePrefix = "arXiv",
    doi = "10.1103/PhysRevC.72.064901",
    journal = "Phys. Rev. C",
    volume = "72",
    pages = "064901",
    year = "2005"
}

@article{SMASH:2016zqf,
    author = "Weil, J. and others",
    collaboration = "SMASH",
    title = "{Particle production and equilibrium properties within a new hadron transport approach for heavy-ion collisions}",
    eprint = "1606.06642",
    archivePrefix = "arXiv",
    primaryClass = "nucl-th",
    doi = "10.1103/PhysRevC.94.054905",
    journal = "Phys. Rev. C",
    volume = "94",
    number = "5",
    pages = "054905",
    year = "2016"
}

@article{Cassing:2009vt,
    author = "Cassing, W. and Bratkovskaya, E. L.",
    title = "{Parton-Hadron-String Dynamics: an off-shell transport approach for relativistic energies}",
    eprint = "0907.5331",
    archivePrefix = "arXiv",
    primaryClass = "nucl-th",
    doi = "10.1016/j.nuclphysa.2009.09.007",
    journal = "Nucl. Phys. A",
    volume = "831",
    pages = "215--242",
    year = "2009"
}

@article{Chen:2015hfc,
	author = "Chen, Hao-Lei and Fukushima, Kenji and Huang, Xu-Guang and Mameda, Kazuya",
	title = "{Analogy between rotation and density for Dirac fermions in a magnetic field}",
	eprint = "1512.08974",
	archivePrefix = "arXiv",
	primaryClass = "hep-ph",
	doi = "10.1103/PhysRevD.93.104052",
	journal = "Phys. Rev. D",
	volume = "93",
	number = "10",
	pages = "104052",
	year = "2016"
}

@article{Jiang:2016wvv,
	author = "Jiang, Yin and Liao, Jinfeng",
	title = "{Pairing Phase Transitions of Matter under Rotation}",
	eprint = "1606.03808",
	archivePrefix = "arXiv",
	primaryClass = "hep-ph",
	doi = "10.1103/PhysRevLett.117.192302",
	journal = "Phys. Rev. Lett.",
	volume = "117",
	number = "19",
	pages = "192302",
	year = "2016"
}

@article{Ebihara:2016fwa,
	author = "Ebihara, Shu and Fukushima, Kenji and Mameda, Kazuya",
	title = "{Boundary effects and gapped dispersion in rotating fermionic matter}",
	eprint = "1608.00336",
	archivePrefix = "arXiv",
	primaryClass = "hep-ph",
	doi = "10.1016/j.physletb.2016.11.010",
	journal = "Phys. Lett. B",
	volume = "764",
	pages = "94--99",
	year = "2017"
}

@article{Mameda:2015ria,
	author = "Mameda, Kazuya and Yamamoto, Arata",
	title = "{Magnetism and rotation in relativistic field theory}",
	eprint = "1504.05826",
	archivePrefix = "arXiv",
	primaryClass = "hep-th",
	doi = "10.1093/ptep/ptw128",
	journal = "PTEP",
	volume = "2016",
	number = "9",
	pages = "093B05",
	year = "2016"
}

@article{Chernodub_2017,
	title={Interacting fermions in rotation: chiral symmetry restoration, moment of inertia and thermodynamics},
	volume={2017},
	ISSN={1029-8479},
	url={http://dx.doi.org/10.1007/JHEP01(2017)136},
	DOI={10.1007/jhep01(2017)136},
	number={1},
	journal={Journal of High Energy Physics},
	publisher={Springer Science and Business Media LLC},
	author={Chernodub, M. N. and Gongyo, Shinya},
	year={2017},
	month=jan }

@article{Chernodub:2017ref,
	author = "Chernodub, M. N. and Gongyo, Shinya",
	title = "{Effects of rotation and boundaries on chiral symmetry breaking of relativistic fermions}",
	eprint = "1702.08266",
	archivePrefix = "arXiv",
	primaryClass = "hep-th",
	doi = "10.1103/PhysRevD.95.096006",
	journal = "Phys. Rev. D",
	volume = "95",
	number = "9",
	pages = "096006",
	year = "2017"
}

@article{Chernodub:2020qah,
	author = "Chernodub, M. N.",
	title = "{Inhomogeneous confining-deconfining phases in rotating plasmas}",
	eprint = "2012.04924",
	archivePrefix = "arXiv",
	primaryClass = "hep-ph",
	doi = "10.1103/PhysRevD.103.054027",
	journal = "Phys. Rev. D",
	volume = "103",
	number = "5",
	pages = "054027",
	year = "2021"
}

@article{Wang:2018sur,
	author = "Wang, Xinyang and Wei, Minghua and Li, Zhibin and Huang, Mei",
	title = "{Quark matter under rotation in the NJL model with vector interaction}",
	eprint = "1808.01931",
	archivePrefix = "arXiv",
	primaryClass = "hep-ph",
	doi = "10.1103/PhysRevD.99.016018",
	journal = "Phys. Rev. D",
	volume = "99",
	number = "1",
	pages = "016018",
	year = "2019"
}

@article{WeiMingHua:2020eee,
	author = "(Minghua Wei, Yin Jiang and Mei Huang",
	title = "{Mass splitting of vector mesons and spontaneous spin polarization under rotation *}",
	eprint = "2011.10987",
	archivePrefix = "arXiv",
	primaryClass = "hep-ph",
	doi = "10.1088/1674-1137/ac338e",
	journal = "Chin. Phys. C",
	volume = "46",
	number = "2",
	pages = "024102",
	year = "2022"
}

@article{Sun:2021hxo,
	author = "Sun, Fei and Huang, Anping",
	title = "{Properties of strange quark matter under strong rotation}",
	eprint = "2104.14382",
	archivePrefix = "arXiv",
	primaryClass = "hep-ph",
	doi = "10.1103/PhysRevD.106.076007",
	journal = "Phys. Rev. D",
	volume = "106",
	number = "7",
	pages = "076007",
	year = "2022"
}

@article{Xu:2022hql,
	author = "Xu, Kun and Lin, Fan and Huang, Anping and Huang, Mei",
	title = "{\ensuremath{\Lambda}/\ensuremath{\Lambda}\textasciimacron{} polarization and splitting induced by rotation and magnetic field}",
	eprint = "2205.02420",
	archivePrefix = "arXiv",
	primaryClass = "hep-ph",
	doi = "10.1103/PhysRevD.106.L071502",
	journal = "Phys. Rev. D",
	volume = "106",
	number = "7",
	pages = "L071502",
	year = "2022"
}

@article{Sun:2023kuu,
	author = "Sun, Fei and Xu, Kun and Huang, Mei",
	title = "{Splitting of chiral and deconfinement phase transitions induced by rotation}",
	eprint = "2307.14402",
	archivePrefix = "arXiv",
	primaryClass = "hep-ph",
	doi = "10.1103/PhysRevD.108.096007",
	journal = "Phys. Rev. D",
	volume = "108",
	number = "9",
	pages = "096007",
	year = "2023"
}

@article{Fujimoto:2021xix,
	author = "Fujimoto, Yuki and Fukushima, Kenji and Hidaka, Yoshimasa",
	title = "{Deconfining Phase Boundary of Rapidly Rotating Hot and Dense Matter and Analysis of Moment of Inertia}",
	eprint = "2101.09173",
	archivePrefix = "arXiv",
	primaryClass = "hep-ph",
	reportNumber = "KEK-TH-2290, J-PARC-TH-0236, RIKEN-iTHEMS-Report-21",
	doi = "10.1016/j.physletb.2021.136184",
	journal = "Phys. Lett. B",
	volume = "816",
	pages = "136184",
	year = "2021"
}

@article{Chernodub:2017mvp,
	author = "Chernodub, M. N. and Gongyo, Shinya",
	title = "{Edge states and thermodynamics of rotating relativistic fermions under magnetic field}",
	eprint = "1706.08448",
	archivePrefix = "arXiv",
	primaryClass = "hep-th",
	doi = "10.1103/PhysRevD.96.096014",
	journal = "Phys. Rev. D",
	volume = "96",
	number = "9",
	pages = "096014",
	year = "2017"
}

@article{Mukherjee:2023ijv,
    author = "Mukherjee, Gaurav and Dutta, Dipanwita and Mishra, Dipak Kumar",
    title = "{Conserved number fluctuations under global rotation in a hadron resonance gas model}",
    eprint = "2304.14658",
    archivePrefix = "arXiv",
    primaryClass = "hep-ph",
    doi = "10.1140/epjc/s10052-024-12592-1",
    journal = "Eur. Phys. J. C",
    volume = "84",
    number = "3",
    pages = "258",
    year = "2024"
}

@article{Mukherjee:2023qvq,
    author = "Mukherjee, Gaurav and Dutta, D. and Mishra, D. K.",
    title = "{An augmented QCD phase portrait: Mapping quark-hadron deconfinement for hot, dense, rotating matter under magnetic field}",
    eprint = "2304.12643",
    archivePrefix = "arXiv",
    primaryClass = "hep-ph",
    doi = "10.1016/j.physletb.2023.138228",
    journal = "Phys. Lett. B",
    volume = "846",
    pages = "138228",
    year = "2023"
}

@article{Sahoo:2026lrw,
    author = "Sahoo, Bhagyarathi and Singh, Captain R. and Sahoo, Raghunath",
    title = "{Probing rotational dynamics of quark gluon plasma via global vorticity}",
    eprint = "2602.13618",
    archivePrefix = "arXiv",
    primaryClass = "hep-ph",
    doi = "10.1016/j.physletb.2026.140714",
    journal = "Phys. Lett. B",
    volume = "880",
    pages = "140714",
    year = "2026"
}

@article{Sahoo:2025fif,
    author = "Sahoo, Bhagyarathi and Pradhan, Kshitish Kumar and Sahu, Dushmanta and Sahoo, Raghunath",
    title = "{Rotational susceptibility of a hot and dense hadronic matter}",
    eprint = "2507.03708",
    archivePrefix = "arXiv",
    primaryClass = "hep-ph",
    doi = "10.1140/epja/s10050-026-01938-w",
    journal = "Eur. Phys. J. A",
    volume = "62",
    number = "8",
    pages = "162",
    year = "2026"
}

@article{Pradhan:2023rvf,
    author = "Pradhan, Kshitish Kumar and Sahoo, Bhagyarathi and Sahu, Dushmanta and Sahoo, Raghunath",
    title = "{Thermodynamics of a rotating hadron resonance gas with van der Waals interaction}",
    eprint = "2304.05190",
    archivePrefix = "arXiv",
    primaryClass = "hep-ph",
    doi = "10.1140/epjc/s10052-024-13283-7",
    journal = "Eur. Phys. J. C",
    volume = "84",
    number = "9",
    pages = "936",
    year = "2024"
}

@article{Sahoo:2023xnu,
    author = "Sahoo, Bhagyarathi and Singh, Captain R. and Sahu, Dushmanta and Sahoo, Raghunath and Alam, Jan-e",
    title = "{Impact of vorticity and viscosity on the hydrodynamic evolution of hot QCD medium}",
    eprint = "2302.07668",
    archivePrefix = "arXiv",
    primaryClass = "hep-ph",
    doi = "10.1140/epjc/s10052-023-12027-3",
    journal = "Eur. Phys. J. C",
    volume = "83",
    number = "9",
    pages = "873",
    year = "2023"
}

@article{Padhan:2026mwg,
    author = "Padhan, Nandita and Pradhan, Kshitish Kumar and Chatterjee, Arghya and Sahoo, Raghunath",
    title = "{Vorticity-induced modifications of chemical freeze-out in heavy-ion collisions}",
    eprint = "2603.27267",
    archivePrefix = "arXiv",
    primaryClass = "hep-ph",
    doi = "10.1016/j.physletb.2026.140604",
    journal = "Phys. Lett. B",
    volume = "879",
    pages = "140604",
    year = "2026"
}

@article{Pradhan:2025pol,
    author = "Pradhan, Kshitish Kumar and Sahu, Dushmanta and Sahoo, Raghunath",
    title = "{Emergent spin polarization from {\ensuremath{\rho}} meson condensation in rotating hadronic matter}",
    eprint = "2510.22755",
    archivePrefix = "arXiv",
    primaryClass = "hep-ph",
    doi = "10.1016/j.physletb.2025.140090",
    journal = "Phys. Lett. B",
    volume = "872",
    pages = "140090",
    year = "2026"
}

@article{Padhan:2025qhz,
    author = "Padhan, Nandita and Chatterjee, Arghya",
    title = "{Thermal conductivity of rotating hot and dense hadronic matter under the influence of Coriolis force}",
    doi = "10.1088/1361-6471/ae2040",
    journal = "J. Phys. G",
    volume = "52",
    number = "12",
    pages = "125101",
    year = "2025"
}

@article{Dwibedi:2025boz,
    author = "Dwibedi, Ashutosh and Marattukalam, Dani Rose J. and Padhan, Nandita and Sahu, Dushmanta and Dey, Jayanta and Goswami, Kangkan and Chatterjee, Arghya and Ghosh, Sabyasachi and Sahoo, Raghunath",
    title = "{Shear viscosity and electrical conductivity of a rotating nuclear medium in the hadron-resonance-gas and Nambu{\textendash}Jona-Lasinio models}",
    eprint = "2505.03588",
    archivePrefix = "arXiv",
    primaryClass = "nucl-th",
    doi = "10.1103/r4cq-stt7",
    journal = "Phys. Rev. C",
    volume = "113",
    number = "4",
    pages = "044903",
    year = "2026"
}

@article{Dwibedi:2024amt,
    author = "Dwibedi, Ashutosh and Padhan, Nandita and Marattukalam, Dani Rose J. and Chatterjee, Arghya and De, Sudipan and Ghosh, Sabyasachi",
    title = "{Effect of coriolis force on diffusion of D meson}",
    eprint = "2411.09983",
    archivePrefix = "arXiv",
    primaryClass = "hep-ph",
    doi = "10.1088/1361-6471/adf983",
    journal = "J. Phys. G",
    volume = "52",
    number = "9",
    pages = "095101",
    year = "2025"
}

@article{Padhan:2024edf,
    author = "Padhan, Nandita and Dwibedi, Ashutosh and Chatterjee, Arghya and Ghosh, Sabyasachi",
    title = "{Effect of Coriolis force on electrical conductivity tensor for the rotating hadron resonance gas}",
    eprint = "2403.16647",
    archivePrefix = "arXiv",
    primaryClass = "hep-ph",
    doi = "10.1103/PhysRevC.110.024904",
    journal = "Phys. Rev. C",
    volume = "110",
    number = "2",
    pages = "024904",
    year = "2024"
}

@article{Aung:2023pjf,
    author = "Aung, Cho Win and Dwibedi, Ashutosh and Dey, Jayanta and Ghosh, Sabyasachi",
    title = "{Effect of Coriolis force on the shear viscosity of quark matter: A nonrelativistic description}",
    eprint = "2303.16462",
    archivePrefix = "arXiv",
    primaryClass = "nucl-th",
    doi = "10.1103/PhysRevC.109.034913",
    journal = "Phys. Rev. C",
    volume = "109",
    number = "3",
    pages = "034913",
    year = "2024"
}

@article{Dwibedi:2023akm,
    author = "Dwibedi, Ashutosh and Aung, Cho Win and Dey, Jayanta and Ghosh, Sabyasachi",
    title = "{Effect of the Coriolis force on the electrical conductivity of quark matter: A nonrelativistic description}",
    eprint = "2305.10183",
    archivePrefix = "arXiv",
    primaryClass = "nucl-th",
    doi = "10.1103/PhysRevC.109.034914",
    journal = "Phys. Rev. C",
    volume = "109",
    number = "3",
    pages = "034914",
    year = "2024"
}

@article{Kumar:2026dtz,
    author = "Kumar, Ankit and Gaur, Diwakar and Chandra, Vinod",
    title = "{Rotating magnetized pion gas of finite transverse size: condensation constraints and transport properties}",
    eprint = "2606.20530",
    archivePrefix = "arXiv",
    primaryClass = "hep-ph",
    month = "6",
    year = "2026"
}

@article{Padhan:2026oyt,
    author = "Padhan, Nandita and Pradhan, Kshitish Kumar and Chatterjee, Arghya and Sahoo, Raghunath",
    title = "{Rotational effects on the chiral crossover transition in QCD matter}",
    eprint = "2609.15513",
    archivePrefix = "arXiv",
    primaryClass = "hep-ph",
    month = "9",
    year = "2026"
}
\end{document}